\documentclass[12pt]{article}
\usepackage{amsmath,amssymb,bm,graphicx,algorithm,algorithmic}
\usepackage[top=25truemm,bottom=20truemm,left=25truemm,right=25truemm]{geometry}
\allowdisplaybreaks

\title{Generalization of generative model for neuronal ensemble inference method}
\author{Shun Kimura\thanks{Department of Mechanics Systems Engineering, Graduate School of Science and Engineering, Ibaraki University, Hitachi, Ibaraki, Japan} \and Koujin Takeda$^*$\thanks{koujin.takeda.kt@vc.ibaraki.ac.jp}}

\date{27th June, 2023}

\begin{document}
\maketitle

\begin{abstract}

Various brain functions that are necessary to maintain life activities materialize through the interaction of countless neurons. 
Therefore, it is important to analyze functional neuronal network.
To elucidate the mechanism of brain function, many studies are being actively conducted on functional neuronal ensemble and hub, including all areas of neuroscience. 
In addition, recent study suggests that the existence of functional neuronal ensembles and hubs contributes to the efficiency of information processing. 
For these reasons, there is a demand for methods to infer functional neuronal ensembles from neuronal activity data, and methods based on Bayesian inference have been proposed. 
However, there is a problem in modeling the activity in Bayesian inference. 
The features of each neuron's activity have non-stationarity depending on physiological experimental conditions. 
As a result, the assumption of stationarity in Bayesian inference model impedes inference, which leads to destabilization of inference results and degradation of inference accuracy. 
In this study, we extend the range of the variable for expressing the neuronal state, and generalize the likelihood of the model for extended variables. By comparing with the previous study, our model can express the neuronal state in larger space. This generalization without restriction of the binary input enables us to perform soft clustering and apply the method to non-stationary neuroactivity data.
In addition, for the effectiveness of the method, we apply the developed method to multiple synthetic fluorescence data generated from
 the electrical potential data in leaky integrated-and-fire model.
\end{abstract}
\vspace{3mm}
\\

\clearpage

\section*{Introduction}
\label{chap:intro}

In many organisms, various functions necessary to maintain life activities such as perception, movement, and emotion are realized through the interaction of countless neurons in multiple regions of brain. Each neuron transmits information by chemical and electrical signals through synaptic and gap junctions. Therefore, from the perspective of network science, analysis of neuronal network is important for elucidating the mechanisms of brain functions. As such, in many fields of neuroscience, various studies are being conducted to elucidate the structure and function of neuronal networks from experimental and theoretical viewpoints. In fact, in the field of neurophysiology, measurement techniques have been developed for the purpose of simultaneously visualizing the activity of neurons in multiple brain regions. One of the main methods for measuring neuronal activity at the level of a single neuron is calcium imaging, which can measure the activity of a large number of neurons on the $10^4$ order at a time\cite{ota2021fast,yu2021diesel2p}. 

In our study, we define functional connections as virtual connections, which are determined based on the synchronization of activities among neurons, as opposed to structural connections representing actual synaptic or gap junction.
Of course, structural connection is significant for clarifying structural neuronal network in the brain. However, it is difficult to infer structural connection from activity data because unobserved neurons may influence the activity of observed neurons in measurement experiment through hidden structural connections.
 For this reason, the study on the inference of functional connectivity from activity data has mainly and intensively been conducted.
For example, the functional connection strength is evaluated as time-averaged quantity
in the methods in the previous studies \cite{mezard2011exact,roudi2011mean,terada2020inferring}.
 
In addition to the functional connection, functional neuronal ensembles are defined as groups of neurons with 
virtual functional connections determined by the synchronization of their activities.
It should be emphasized that the synchronous activities of neurons do not necessarily depend on the presence or absence of 
direct structural coupling, but may also on indirect structural coupling through multiple neurons.
Such functional neuronal ensembles play a specific role in the information processing in the brain, and 
various methods for inferring functional neuronal ensembles have been proposed \cite{lopes2011neuronal,lopes2013detecting,molter2018detecting}.
It can also be used for limiting the data to be analyzed as preprocessing of local functional network inference.  
As another significant topic, it has been suggested that hub neurons, which have functional connections with many neurons and influence them in neuronal network, are involved in the efficiency of the information processing between neurons \cite{gal2021role}. 
Thus, inference of functional neuronal ensembles centered on hub neurons
is important not only as a preprocessing for local functional network inference but also for elucidating the global mechanism of brain functions.

Among the methods of functional neuronal ensemble, Bayesian inference model and Markov chain Monte Carlo (MCMC) method are used in the previous study\cite{diana2019bayesian}, whose improvement for faster convergence to inference result has also been proposed \cite{kimura2021improved}.
The method based on Bayesian inference needs a generative model, where only binary expression of neuronal activity data is allowed in the previous studies. For this reason, we call the method in the previous studies \cite{diana2019bayesian, kimura2021improved}
as {\it binary inference method} (BIM) in the following.
Apparently, BIM cannot be applied to activity data of continuous values such as fluorescence intensity by calcium imaging. Of course, it is possible to apply BIM to continuous-valued data after binarization. However, it is desirable to analyze continuous-valued data directly, because part of the original information in raw activity data may be lost by binarization. Another problem in BIM is the assumption of stationarity in the activity, which means that functional neuronal ensemble is assumed to be unchanged throughout the measurement experiment. However, functional neuronal ensembles will change depending on the brain state and experimental task.
This is because actual neurons often have multiple functional roles depending on brain states such as sleep and wakefulness. Furthermore, responses to stimuli are different due to various brain states.
Therefore, when BIM is applied to highly non-stationary activity data with frequent changes of brain states \cite{miyamoto2016top,manita2015top}, the inference result by MCMC may be unstable due to the dependence on initial condition in MCMC,
 and accuracy of the inference may worsen. Additionally, such disadvantages are also due to hard clustering in BIM, where each neuron must 
belong to a particular ensemble at any time of experiment. Because functional ensembles change over time in non-stationary data, hard clustering method is not suitable.

For these reasons, we develop a generalized method for functional ensemble inference in this study, which does not depend on the format of the neuronal activity data and can be implemented as a soft clustering method, where each neuron may belong to multiple ensembles.
In our generative model, the variable representing the input neuronal activity 
is continuous, and accordingly the limitation of the binary data format is removed.
In addition, by expressing the neuron assignment to ensemble as continuous-valued weight vector rather than categorical variable, 
we can change the hard clustering in BIM to soft one. This can also be regarded as a generalization 
of the model because soft clustering includes hard one.
The generalizations to continuous input activity and soft clustering allow us to widen the range of variables representing neuronal activity in the model. In other words, we can enhance the expressivity of our generative model for neuronal activity,
where expressivity means the range of the variables in the inference model in mathematical sense.
Furthermore, we also develop an algorithm using MCMC for our generalized model. 

For validity, we apply our proposed method to synthetic fluorescence intensity data.
For synthetic data generation, we first generate time series data of electrical membrane potential in neurons under external input current 
by leaky integrated-and-fire model\cite{gerstner2014neuronal}. 
Next, membrane potential is converted to fluorescence intensity by considering the measurement process of experimental calcium imaging. 
After fluorescence intensity data is obtained, we apply our proposed method to infer functional neuronal ensembles.
Finally, we discuss the results in several cases of external input currents 
by tuning the parameter of soft/hard clustering, and compare the results by our proposed method and BIM. 

\section*{Materials and methods}

\subsection*{Bayesian inference model}
\label{chap:model}

The notation of variables in this article is based on the previous work \cite{diana2019bayesian}.
The differences in the notation from the previous study 
due to the generalization of the model will be summarized later.
Boldface is used to denote a vector or matrix variable unless otherwise specified. 
Additionally, subscript is used to denote each element in a vector or matrix variable. 

In our model, let $N$ be the number of neurons, $M$ be the number of time steps in activity measurement, and $A$ be the number of neuronal ensembles.
The generative model in our proposed Bayesian inference method is expressed using three variables: $\bm s$ is neuronal activity; $\bm t$ is the weight of a neuron belonging to a neuronal ensemble; and $\bm \omega$ is time-series activity state of a neuronal ensemble.
The variable $\bm s$ representing neuronal activity is given as input. Accordingly, 
the number of neurons $N$ and the number of time steps $M$ are determined from the input.
In BIM and our proposed method, the number of neuronal ensembles is changed dynamically.
Hence, inferences of ensemble structure and the number of ensembles are performed simultaneously from given activity data. 
Therefore, the number of neuronal ensembles $A$ is an estimator, which is a natural number between $1$ and $N$ 
for hard clustering in BIM.
However, due to the extension of our proposed method to soft clustering, $A$ may be larger than 
the number of neurons $N$, and the possible value of $A$ is in the whole natural number.
In practice, the number of ensembles is expected to be sufficiently smaller than $N$ 
if an appropriate ensemble structure is estimated, therefore our proposed method provides an upper bound 
on $A$ as a parameter.
The estimation of $A$ will be described again in the description of the algorithm.
Next, the subscripts $i$, $k$, and $\mu$ used in elementwise notation of each variable represent the label of the neuron, 
the label of the time step, and the label of the neuronal ensemble, respectively. Therefore, each variable represents the following. 

\begin{itemize}
\item
$s_{i k}$: neuronal activity of the $i$th neuron at the $k$th time step

\item
$t_{\mu i}$: membership weight of the $i$th neuron to the $\mu$th ensemble

\item
$\omega_{k \mu}$: activity of the $\mu$th ensemble at the $k$th time step
\end{itemize}
The ranges of variables ${\bm s}$ and ${\bm \omega}$ are limited as $0 \leq  s_{i k}  \leq 1$ $\forall i, k$ and $0 \leq  \omega_{k \mu}  \leq 1$ $\forall k, \mu$, respectively, to normalize the activity weights of each neuron and ensemble.
The variable ${\bm t}_i = \left\{ t_{1i}, \cdots , t_{Ai} \right\} $ is one-hot representation or $t_{\mu i} \in \{ 0,1 \}$ in the case of hard clustering, while $t_{\mu i}$ ranges from 0 to 1 and satisfies $\sum_{\mu = 1}^{A} t_{\mu i} = 1 \ \ \forall i$ in the case of soft clustering.
In comparison with BIM, the input neural activity $\bm s$ and ensemble activity $\bm \omega$ are changed 
from $0$-or-$1$ binary variables to continuous ones in the range $[0,1]$, and the weight $\bm t$ of neurons is changed 
from categorical variables to continuous-valued vector variables.
The range $[0,1]$ for $\bm s$ and $\bm \omega$ is designed to normalize the amount of information in the activity of each neuron or ensemble.
In addition, the number of ensembles initially given in MCMC is used as the upper bound of $A$, which is denoted by $A_{\rm init}$ in
Table \ref{tab:VarDef}.
The estimation of the appropriate $A$ and ensemble structure under given upper bound  
are common both in BIM and our proposed model.
The detail of the variation in the number of ensembles will be explained later in the algorithm part.
The specific changes in variables are summarized in Table \ref{tab:VarDef}.
\begin{table}
\begin{center}
\caption{Variables in BIM and our proposed method}
\label{tab:VarDef}
\begin{scriptsize}
\begin{tabular}{c||c|c||c|c}  
& what to describe
& input/inferred
& BIM
& our proposed method \\ \hline \hline
  ${\bm s}$ 
& neuronal activity
& input
& $s_{i k} \in \{ 0, 1 \}$ 
& $0 \leq s_{i k} \leq 1$ \\
  ${\bm t}$
& membership weight
& to be inferred
& $t_{i} \in \{ 1, \cdots , A \}$ 
& $0 \leq t_{\mu i} \leq 1 \left( \sum_{\mu = 1}^{A} t_{\mu i} = 1 \right)$ \\
  ${\bm \omega}$
& ensemble activity
& to be inferred
& ${\omega}_{k \mu} \in \{ 0, 1 \}$ 
& $0 \leq {\omega}_{k \mu} \leq 1$ \\ 
  $N$ 
& the number of neurons
& input
& $N \in \mathbb{N}$
& $N \in \mathbb{N}$ \\ 
  $M$ 
& the number of time steps
& input
& $M \in \mathbb{N}$
& $M \in \mathbb{N}$ \\
  $A$
& the number of ensembles
& to be inferred
& $1 \leq A \leq N, \left( A \in \mathbb{N} \right)$
& $1 \leq A \leq A_{{\rm init}} \left( A \in \mathbb{N} \right)$ \\
\hline
\end{tabular}
\end{scriptsize}
\end{center}
\end{table}

In addition to these variables, three parameters ${\bm n}, {\bm p}$, and ${\bm \lambda}$ are introduced to each ensemble in the generative model, whose meanings are described in the following. For each parameter, Dirichlet or beta distribution is assumed as prior distribution to facilitate marginalization in Bayesian inference.

\begin{itemize}
\item
${\bm n}$: 
affinity parameter to a neuronal ensemble, following Dirichlet distribution
\item
${\bm p}$: 
activity parameter of a neuronal ensemble, following beta distribution
\item
${\bm \lambda}$: 
synchronization parameter between the activities of a neuronal ensemble and neurons within the ensemble, following beta distribution
\end{itemize}
More precisely, two kinds of $\bm \lambda$'s  are introduced for synchronization; ${\bm \lambda_1}$ or ${\bm \lambda_0}$ is the synchronization when both of ensemble and neuron are active or inactive, respectively. The distribution of parameters in this model is expressed as follows.
\begin{eqnarray}
P(n_1, \cdots, n_A  |  \alpha_{1}^{(n)}, \cdots, \alpha_{A}^{(n)}) 
& = & {\rm Dir} \left( \alpha_{1}^{(n)}, \cdots, \alpha_{A}^{(n)} \right)
\label{eq:prior_n}
 \propto  \prod_{\mu = 1}^{A} n_{\mu}^{\alpha_{\mu}^{(n)}-1}, \ \ 
\\
P(p_{\mu}  |  \alpha_{\mu}^{(p)}, \beta_{\mu}^{(p)}) 
& = & {\rm Beta} 
      \left( \alpha_{\mu}^{(p)}, \beta_{\mu}^{(p)} \right) 
\label{eq:prior_p} 
 \propto  p_{\mu}^{\alpha_{\mu}^{(p)} - 1} \left( 1 - p \right)^{\beta_{\mu}^{(p)} - 1}, \ \
\\
P({\lambda}_{1, \mu}  |  \alpha_{1, \mu}^{(\lambda)}, \beta_{1, \mu}^{(\lambda)}) 
& = & {\rm Beta} 
      \left( \alpha_{1, \mu}^{(\lambda)}, 
             \beta_{1, \mu}^{(\lambda)} \right)
\label{eq:prior_lambda1} 
 \propto  \lambda_{1, \mu}^{\alpha_{1, \mu}^{(\lambda)} - 1} 
            \left( 1 - \lambda_{1, \mu} \right)^{\beta_{1, \mu}^{(\lambda)} - 1}, \ \
\\
P({\lambda}_{0, \mu}  |  \alpha_{0, \mu}^{(\lambda)}, \beta_{0, \mu}^{(\lambda)}) 
& = & {\rm Beta} 
      \left( \alpha_{0, \mu}^{(\lambda)}, 
             \beta_{0, \mu}^{(\lambda)} \right)
\label{eq:prior_lambda0} 
 \propto  \lambda_{0, \mu}^{\alpha_{0, \mu}^{(\lambda)} - 1} 
            \left( 1 - \lambda_{0, \mu} \right)^{\beta_{0, \mu}^{(\lambda)} - 1}. \ \
\end{eqnarray}
Here $\alpha_{\mu}^{(n)}$ is the hyperparameter of the prior Dirichlet distribution on neuron assignment, which
represents the frequency of neuron assignment to each ensemble. 
$\alpha_{\mu}^{(p)}$ and $\beta_{\mu}^{(p)}$ are the hyperparameters of the prior beta distribution, which represent the frequency of ensemble activity and inactivity, respectively. 
$\alpha_{1,\mu}^{(\lambda)}$ and $\beta_{1,\mu}^{(\lambda)}$ are the hyperparameters of the prior beta distribution, which represent the frequency of active and inactive neurons when the belonging ensemble is active, respectively.
$\alpha_{0,\mu}^{(\lambda)}$ and $\beta_{0,\mu}^{(\lambda)}$ are the hyperparameters of the prior beta distribution, which represent the frequency of active and inactive neurons when the belonging ensemble is inactive, respectively.

The likelihood is designed as follows to facilitate marginalization like the design of prior distribution.
\begin{eqnarray}
P({\bm t}, {\bm \omega}, {\bm s}
| {\bm n}, {\bm p}, {\bm \lambda} )
& \propto & 
  \left( 
    \prod_{\mu = 1}^{A} 
    n_{\mu}^{\tilde{\alpha}_{\mu}^{(n)}}
  \right)
  \cdot
  \left( 
    \prod_{\mu = 1}^{A}  
    p_{\mu}^{\tilde{\alpha}_{\mu}^{(p)}} 
    (1 - p_{\mu})^{\tilde{\beta}_{\mu}^{(p)}}
  \right)
\nonumber \\
& & \cdot 
  \left(
    \prod_{\mu = 1}^{A}
    \left[ 
      \lambda_{1, \mu}
      ^{\tilde{\alpha}_{1, \mu}^{(\lambda)}} 
      \left( 1 - \lambda_{1, \mu} \right)
      ^{\tilde{\beta}_{1, \mu}^{(\lambda)}}
    \right]
  \right) 
\cdot  
  \left(
    \prod_{\mu = 1}^{A}
    \left[ 
      \lambda_{0, \mu}
      ^{\tilde{\alpha}_{0, \mu}^{(\lambda)}} 
      \left( 1 - \lambda_{0, \mu} \right)
      ^{\tilde{\beta}_{0, \mu}^{(\lambda)}}
    \right]
  \right), \nonumber \\
\end{eqnarray}
where
\begin{eqnarray}
\tilde{\alpha}_{\mu}^{(n)} & = & \sum_{i = 1}^{N} t_{\mu i},
\label{eq:tildealpha} \\
\tilde{\alpha}_{\mu}^{(p)} & = & \sum_{k = 1}^{M} \omega_{k \mu},
\\
\tilde{\beta}_{\mu}^{(p)} & = & \sum_{k = 1}^{M} \left( 1 - \omega_{k \mu} \right),
\\
\tilde{\alpha}_{1, \mu}^{(\lambda)} 
& = & \sum_{k = 1}^{M} \omega_{k \mu} 
      \sum_{i = 1}^{N} t_{\mu i} s_{i k},
\\
\tilde{\beta}_{1, \mu}^{(\lambda)} 
& = & \sum_{k = 1}^{M} \omega_{k \mu} 
      \sum_{i = 1}^{N} t_{\mu i} \left(1 - s_{i k} \right),
\\
\tilde{\alpha}_{0, \mu}^{(\lambda)} 
& = & \sum_{k = 1}^{M} \left( 1 - \omega_{k \mu} \right) 
      \sum_{i = 1}^{N} t_{\mu i} s_{i k},
\\
\tilde{\beta}_{0, \mu}^{(\lambda)} 
& = & \sum_{k = 1}^{M} \left( 1 - \omega_{k \mu} \right) 
      \sum_{i = 1}^{N} t_{\mu i} \left(1 - s_{i k} \right).
\label{eq:tildebeta}
\end{eqnarray}
The meanings of these quantities are the same as the hyperparameters, which have already been defined without tildes.
Note that they are defined by the variables $\bm t, \bm \omega$, and $\bm s$. For inference, we need the joint distribution of $\bm t, \bm \omega$, and $\bm s$ under given hyperparameters.
In BIM, the above likelihood was expressed in the form of Kronecker delta, since $\bm t, \bm \omega$, and $\bm s$ take discrete values. The correspondence between the variables in BIM and ours is shown in Table \ref{tab:LikeDef}.
As can be seen from the table, the likelihood in our model is the generalization to continuous variables, which 
also includes binary and categorical cases.
\begin{table}
\begin{center}
\caption{Parameters in the likelihood: The second column from the left is for corresponding hyperparameter without tilde, 
namely what distribution the hyperparameter describes and the meaning of hyperparameter as mentioned in the text.}
\label{tab:LikeDef}
\begin{scriptsize}
\begin{tabular}{c||c||c|c}
& distribution and meaning
& BIM
& our proposed method \\
& of hyperparameter
&
&
\\ \hline \hline
  $\tilde{\alpha}_{\mu}^{(n)}$ 
& Dirichlet,
& $\sum_{i = 1}^{N} \delta_{\mu , t_{i}}$ 
& $\sum_{i = 1}^{N} t_{\mu i}$ \\
  & assignment of neuron & & \\
  $\tilde{\alpha}_{\mu}^{(p)}$
& beta, 
& $\sum_{k = 1}^{M} \delta_{1, \omega_{k \mu}}$ 
& $\sum_{k = 1}^{M} \omega_{k \mu}$ \\
  &for ensemble activity  & & \\
  $\tilde{\beta}_{\mu}^{(p)}$
& beta,
& $\sum_{k = 1}^{M} \delta_{0, \omega_{k \mu}}$ 
& $\sum_{k = 1}^{M} \left( 1 - \omega_{k \mu} \right)$ \\
  &for ensemble inactivity  & & \\
  $\tilde{\alpha}_{1, \mu}^{(\lambda)}$
& beta,
& $\sum_{k = 1}^{M} \delta_{1, \omega_{k \mu}} 
   \sum_{i = 1}^{N} \delta_{\mu , t_{i}} \delta_{1 , s_{i k}} $
& $\sum_{k = 1}^{M} \omega_{k \mu}  
   \sum_{i = 1}^{N} t_{\mu i} s_{i k}$ \\
  & active neurons & & \\
  & under active ensemble & & \\
  $\tilde{\beta}_{1, \mu}^{(\lambda)}$
& beta,
& $\sum_{k = 1}^{M} \delta_{1, \omega_{k \mu}} 
   \sum_{i = 1}^{N} \delta_{\mu , t_{i}} \delta_{0 , s_{i k}} $
& $\sum_{k = 1}^{M} \omega_{k \mu}  
   \sum_{i = 1}^{N} t_{\mu i} \left(1 - s_{i k} \right)$ \\
  & inactive neurons & & \\
  & under active ensemble & & \\
  $\tilde{\alpha}_{0, \mu}^{(\lambda)}$
& beta,
& $\sum_{k = 1}^{M} \delta_{0, \omega_{k \mu}} 
   \sum_{i = 1}^{N} \delta_{\mu , t_{i}} \delta_{1 , s_{i k}} $
& $\sum_{k = 1}^{M} \left( 1 - \omega_{k \mu} \right) 
   \sum_{i = 1}^{N} t_{\mu i} s_{i k}$ \\
  & active neurons & & \\
  & under inactive ensemble & & \\
  $\tilde{\beta}_{0, \mu}^{(\lambda)}$
& beta,
& $\sum_{k = 1}^{M} \delta_{0, \omega_{k \mu}} 
   \sum_{i = 1}^{N} \delta_{\mu , t_{i}} \delta_{0 , s_{i k}} $
& $\sum_{k = 1}^{M} \left( 1 - \omega_{k \mu} \right) 
   \sum_{i = 1}^{N} t_{\mu i} \left(1 - s_{i k} \right)$ \\
  & inactive neurons & & \\
  & under inactive ensemble & & \\
\hline
\end{tabular}
\end{scriptsize}
\end{center}
\end{table}
By product rule in Bayesian statistics and marginalization of the parameters $\bm n, \bm p, \bm \lambda$, the following expression is obtained,
\begin{eqnarray}
& &
P(\bm t, \bm \omega, \bm s | \bm \alpha^{(n)}\!, \bm \alpha^{(p)}\!, \bm \beta^{(p)}\!, \bm \alpha_1^{(\lambda)}\!, \bm \beta_1^{(\lambda)}\!, \bm \alpha_0^{(\lambda)}\!, \bm \beta_0^{(\lambda)} )
\nonumber \\
&=&
 \int  d {\bm n} d {\bm p} d {\bm \lambda}
P(\bm t, \bm \omega, \bm s, \bm n, \bm p, \bm \lambda | \bm \alpha^{(n)}\!, \bm \alpha^{(p)}\!, \bm \beta^{(p)}\!, \bm \alpha_1^{(\lambda)}\!, \bm \beta_1^{(\lambda)}\!, \bm \alpha_0^{(\lambda)}\!, \bm \beta_0^{(\lambda)} )
\nonumber \\
&=&
 \int  d {\bm n} d {\bm p} d {\bm \lambda}
P(\bm t, \bm \omega, \bm s | \bm n, \bm p, \bm \lambda )
P( \bm n, \bm p, \bm \lambda | \bm \alpha^{(n)}\!, \bm \alpha^{(p)}\!, \bm \beta^{(p)}\!, \bm \alpha_1^{(\lambda)}\!, \bm \beta_1^{(\lambda)}\!, \bm \alpha_0^{(\lambda)}\!, \bm \beta_0^{(\lambda)} )
\nonumber \\
&\varpropto & 
\int  d {\bm n} d {\bm p} d {\bm \lambda}
\left(
\prod_{\mu=1}^{A} 
n_{\mu}^{\left\{ \alpha_{\mu}^{(n)} + \tilde{\alpha}_{\mu}^{(n)} \right\} - 1} 
\right)
 \cdot 
\prod_{\mu=1}^{A} 
\left(
p_{\mu}^{\left\{ \alpha^{(p)}_{\mu} + \tilde{\alpha}_{\mu}^{(p)} \right\} - 1}
\left(1 - p_{\mu}\right)^{\left\{ \beta^{(p)}_{\mu} + \tilde{\beta}_{\mu}^{(p)} \right\} - 1}
\right)
\nonumber \\
&& \cdot 
\left(
\lambda_{1, \mu}^{\left\{ \alpha^{(\lambda)}_{1, \mu}
                          + \tilde{\alpha}_{1, \mu}^{(\lambda)} \right\} - 1}
\left(1 - \lambda_{1, \mu} \right)^{\left\{ \beta^{(\lambda)}_{1, \mu} 
                                  + \tilde{\beta}_{1, \mu}^{(\lambda)} \right\} - 1} 
\right)
\nonumber \\
&& \cdot 
\left(
\lambda_{0, \mu}^{\left\{ \alpha^{(\lambda)}_{0,\mu} 
                          + \tilde{\alpha}_{0, \mu}^{(\lambda)} \right\} - 1}
\left(1 - \lambda_{0,\mu} \right)^{\left\{ \beta^{(\lambda)}_{0, \mu} 
                                           + \tilde{\beta}_{0, \mu}^{(\lambda)} \right\} - 1} 
\right)
\nonumber \\
& = &
{\cal B} \left( \alpha_{1}^{(n)} + \tilde{\alpha}_{1}^{(n)}, \cdots,  
                \alpha_{A}^{(n)} + \tilde{\alpha}_{A}^{(n)} 
         \right)
 \prod_{\mu = 1}^{A} 
B \left( \alpha_{\mu}^{(p)} + \tilde{\alpha}_{\mu}^{(p)} ,
         \beta_{\mu}^{(p)} + \tilde{\beta}_{\mu}^{(p)} 
 \right)
\nonumber \\ 
&&B \left( \alpha_{1 , \mu}^{(\lambda)} + \tilde{\alpha}_{1 , \mu}^{(\lambda)} ,
         \beta_{1 , \mu}^{(\lambda)} + \tilde{\beta}_{1 , \mu}^{(\lambda)}
 \right)
B \left( \alpha_{0 , \mu}^{(\lambda)} + \tilde{\alpha}_{0 , \mu}^{(\lambda)} ,
         \beta_{0 , \mu}^{(\lambda)} + \tilde{\beta}_{0 , \mu}^{(\lambda)}
 \right),
\label{eq:mymodel}
\end{eqnarray}
where $B(x_1 , x_2)$ is beta function and ${\cal B} (x_{1} , \cdots x_{A})$ is multivariate beta function represented as
\begin{eqnarray}
B ( x_1 , x_2 )
& = &
\frac{{\it \Gamma} \left( x_1 \right) {\it \Gamma} \left( x_2 \right)}
     {{\it \Gamma} \left( x_1 + x_2 \right)},
\\
{\cal B} (x_1, \cdots, x_A) 
& = &
\frac{\prod_{\mu = 1}^{A} {\it \Gamma} (x_{\mu})}
     {{\it \Gamma} \left( \sum_{\mu = 1}^{A} x_{\mu} \right)},
\end{eqnarray}
with ${\it \Gamma} (x)$ being gamma function.
For inference of the structure of the functional neuronal ensemble and activity of the ensemble, namely inference of $\bm t$ and $\bm \omega$, the marginalized expression (\ref{eq:mymodel}) is used for improvement of inference accuracy and reduction of computational cost.
In more detail, Dirichlet and beta distributions are chosen as the conjugate prior 
both in BIM and in our model so that the forms of the prior and posterior distributions are the same.
In addition, the form of prior distribution is designed to remain unchanged after update of the hyperparameters.
Such design enables us to integrate out the parameters $\bm n, \bm p$, and $\bm \lambda$ analytically as in Eq (\ref{eq:mymodel}),
which yields probability model expressed only by hyperparameters. Consequently,
computational cost and estimation error can be reduced, because the parameters $\bm n,\bm p$ and $\bm \lambda$ can be eliminated by
analytical integration, and accordingly the numerical integration of these parameters is not necessary for the inference.

To summarize, we compare the two models: BIM and ours. 
In our model, the variable range of neuronal state is extended by the continuous $\bm s$ and $\bm \omega$, which 
leads to more detailed representation of the similarity between activities of neurons and ensemble.
Membership weight $\bm t$ is generalized to continuous-valued vector variable for expressing the certainty of 
membership assignment to target ensemble.
By changing $\bm t$ to continuous-valued, our generative model allows soft clustering for expressing overlap 
in membership assignment, and accordingly multiple roles of neurons can also be expressed.
Moreover, the likelihood in BIM is generalized to improve the expressivity of the model.
Consequently, our model is very effective not only for stationary activity data with the fixed functional role of neuron
but also for the data of non-stationary activity, the data with multiple functional roles for each neuron, 
or the data under the experiment of low reproducibility.
Additionally, MCMC algorithm in BIM 
is also changed for estimation of continuous variables. The changes in the algorithm will be described in the following.

\subsection*{MCMC and Dirichlet process}
\label{chap:MCMC}
In our method, the structure of the neuronal ensemble and activity of ensemble are inferred from the solution of posterior maximization by applying MCMC method to Eq (\ref{eq:mymodel}) similarly to BIM \cite{diana2019bayesian}. In addition, the number of neuronal ensembles is dynamically changed by Dirichlet process in MCMC. Dirichlet process is a stochastic process that can generate arbitrary discrete distributions, and the probability of each event is determined by concentration parameter \cite{neal2000markov}. For speeding up the inference by MCMC, parallel computation using synchronous update can be implemented.

The variables to be inferred are ${\bm t}$ and ${\bm \omega}$ in our model, and the variable updates for $\bm t$ and $\bm \omega$ are performed alternately until convergence in MCMC. The procedure of variable update is summarized as follows.

\begin{enumerate}
\item
Ensemble activity ${\bm \omega}$ is updated according to Dirichlet process of order two (or beta process) 
and Metropolis-Hastings method in MCMC.
\label{update1}
\item
Value of membership weight ${\bm t}$ is updated according to Dirichlet process and Metropolis-Hastings method in MCMC.
\label{update2}
\item
Hyperparameters are updated using updated ${\bm \omega}$ and ${\bm t}$ by the processes 1 and 2.
\end{enumerate}

The update rule of membership weight $\bm t_i = \{ t_{1 i}, \ldots, t_{A i} \}$ is explained in the following, which differs from BIM
 due to the change of the generative model.  Other variables are updated in the same manner as in BIM.

In our model, the element of membership weight $t_{\mu i}$ is a continuous variable satisfying
normalization condition $\sum_{\mu =1}^{A} t_{\mu i} = 1$.  
Hence, for update of membership weight $\bm t_i$, all elements in $\bm t_i$ must be updated unlike Dirichlet process for one-hot representation of $\bm t_i$ in BIM.
Let ${\bm t}_{i}^{0}$ be the original membership weight of neuron ${i}$ before update and ${\bm t}_{i}^{*}$ be the proposed membership weight for update, which may or may not be accepted by Metropolis-Hastings method. 
Here we define {\it concentration ensemble} $G_i^*$ for neuron $i$ in each iteration of MCMC.
The values in elements of the proposed membership weights $\bm t^*_i$ will concentrate on the ensemble $G_i^*$ by Dirichlet process with {\it transition parameter} $\alpha^{(t)}$.
More precisely, in Dirichlet process the concentration ensemble $G_i^*$ is determined first, next the weight $t_{G_i^* i}$ is increased by $\alpha^{(t)}$ and the weights of other ensembles remain unchanged, then finally all elements of the proposed weight are normalized so as to satisfy the condition of probability.
In the case of hard clustering as in BIM, such normalization is not necessary  because of one-hot representation of the weight.

The concentration ensemble $G_i^*$ is determined according to the transition probability in Dirichlet process
 as in the following Eq \eqref{eq:TranRate_t}.
The probability for the concentration ensemble $G_i^*$ is proportional to the size of the ensemble $G_i^*$, that is, 
the sum of the proposed weights of the neurons belonging to the ensemble $G_i^*$ \cite{neal2000markov,jain2004split}.
\begin{eqnarray}
\label{eq:TranRate_t}
Q(G_i^{*} | {\bm t}_{\backslash i}^{0}) =
\frac{\sum_{j = 1, j \neq {i}}^{N} t_{G_i^{*} j}^{0}}
     {N - 1},
\end{eqnarray}
where $\bm t^0 = \{ \bm t_{1}^0, \ldots, \bm t_{N}^0 \}$.
 Then, the proposed weights are computed as in Eq \eqref{eq:Tran_t}, 
 which includes the increase of the weight $t_{G^* i}$ and the normalization.
\begin{eqnarray}
\label{eq:Tran_t}
t_{{\mu} {i}}^{*}
=
\left\{
\begin{array}{ll}
\displaystyle \frac{t_{\mu {i}}^{0}}{1 + \alpha^{(t)}} & {\rm if} \ \ {\mu} \neq G_i^*, \vspace{2mm} \\
\displaystyle \frac{t_{\mu {i}}^{0} + \alpha^{(t)}}{1 + \alpha^{(t)}} & {\rm if} \ \ {\mu} = G_i^*.
\end{array}
\right.
\end{eqnarray}

Conversely, to satisfy the detailed balance condition in MCMC, 
the reverse process for the determination of concentration ensemble must be considered.
The probability of concentration ensemble before update, denoted by $G_i^0$, is expressed 
 by Eq (\ref{eq:update11}) under given membership weight after update, namely $\bm t_i^*$. 
\begin{eqnarray}
Q(G_i^0 | {\bm t}_{\backslash i}^{0}) =
\frac{\sum_{j = 1, j \neq i}^{N} t_{G_i^0 j}^{0}}
     {N - 1},
\label{eq:update11}
\end{eqnarray}
where backslash means removal of specific element, $\bm t_{\backslash i}^0 = \{ \bm t_{1}^0, \ldots, \bm t_{i-1}^0, \bm t_{i+1}^0 , \ldots, \bm t_{N}^0 \}$.

Here we should comment on the difference in the inference of membership weight $\bm t$ between BIM and our model
for later convenience.
In BIM, the membership weight $\bm t$ is a categorical variable, therefore the concentration ensemble 
$G_i^*$ is not necessary in determining the transition destination, and
the transition probability to each ensemble is defined in proportion to the size of ensemble.
Since there may be a transition to a new ensemble according to Dirichlet process, 
the cases of transitions to an existing ensemble and to new ensemble should be described separately.
Hence, in BIM, the transition probabilities under the weight $\bm t$ in one-hot representation are given as follows.
\begin{eqnarray}
Q( t_{\mu i}^{*} = 1 | {\bm t}_{\backslash i}^{0}) 
=
\left\{
\begin{array}{ll}
\displaystyle \frac{\sum_{j = 1, j \neq i}^{N} t_{\mu i}^{0}}{N - 1 + \alpha_{\rm new}} 
& {\rm if} \ \ {\mu} \leq A, \vspace{2mm} \\
\displaystyle \frac{\alpha_{\rm new}}{N - 1 + \alpha_{\rm new}} & {\rm if} \ \ {\mu} = A + 1,
\end{array}
\right.
\label{eq:update11sub}
\end{eqnarray}
where $\alpha_{\rm new}$ represents concentration parameter in Dirichlet process,
 and larger value of $\alpha_{\rm new}$ results in larger $A$.
In our model, $\alpha_{\rm new}$ is set to 0 because there is no upper bound for $A$ due to soft clustering.

We go back to the description of our model.
The ratio of probabilities for the concentration ensembles $G^*_i, G^*_0$ is given by 
\begin{eqnarray}
\frac{ Q (G_i^{0} | {\bm t}_{\backslash i}^{0} ) }
     { Q (G_i^{*} | {\bm t}_{\backslash i}^{0} ) }
=
\frac{ \sum_{j = 1, j \neq i}^{N} t_{G_i^{0} j} }
     { \sum_{j = 1, j \neq i}^{N} t_{G_i^{*} j} }.
\label{eq:update12}
\end{eqnarray}
The derived probabilities ratio is represented by the ratio of the ensemble sizes.
Remember that $\bm t_i^*$ is the proposed membership weight, and  
for acceptance of the proposed weight it is necessary to calculate the acceptance rate in Metropolis-Hastings method.
The acceptance rate of the proposed weight $\bm t_i^*$ under the original weight $\bm t_i^0$ is given as 
\begin{eqnarray}
\label{eq:AcceptRate_t}
a \left( {\bm t}_{i}^{*}, {\bm t}_{i}^{0} \right)
 = 
 \min  \left(1,
           \frac{P ( {\bm t_i^*}, {\bm t_{\backslash i}^0}, {\bm \omega}, {\bm s} ) }
                {P ( {\bm t^0}, {\bm \omega}, {\bm s} ) }
           \frac{ Q (G_i^{0} | {\bm t}_{\backslash i}^{0}) }
                { Q (G_i^{*} | {\bm t}_{\backslash i}^{0}) }
    \right),
\end{eqnarray}
where hyperparameters in $P$ are omitted and
\begin{eqnarray}
& \ \ &
\frac{P ( {\bm t_i^*}, {\bm t_{\backslash i}^0}, {\bm \omega}, {\bm s} ) }
     {P ( {\bm t^0}, {\bm \omega}, {\bm s} ) }
\frac{ Q (G_i^{0} | {\bm t}_{\backslash i}^{0}) }
     { Q (G_i^{*} | {\bm t_{\backslash i}^{0}}) }  \nonumber \\
& = & 
\frac{ {\cal B} \left( \alpha_{1}^{(n)} + \tilde{\alpha}_{1}^{(n)}, \cdots,  
                       \alpha_{A}^{(n)} + \tilde{\alpha}_{A}^{(n)} 
               \right)
       \Big| _{{\bm t}_{i} = {\bm t}_{i}^{*}}        
     }
     { {\cal B} \left( \alpha_{1}^{(n)} + \tilde{\alpha}_{1}^{(n)}, \cdots,  
                       \alpha_{A}^{(n)} + \tilde{\alpha}_{A}^{(n)} 
               \right)
       \Big| _{{\bm t}_{i} = {\bm t}_{i}^{0}}        
     }
\nonumber \\
& \ \ & \cdot         
\frac{         
       \prod_{\mu = 1}^{A}      
	   B \left( \alpha_{1 , \mu}^{(\lambda)} + \tilde{\alpha}_{1 , \mu}^{(\lambda)} ,
                \beta_{1 , \mu}^{(\lambda)} + \tilde{\beta}_{1 , \mu}^{(\lambda)}
        \right)
       B \left( \alpha_{0 , \mu}^{(\lambda)} + \tilde{\alpha}_{0 , \mu}^{(\lambda)} ,
                \beta_{0 , \mu}^{(\lambda)} + \tilde{\beta}_{0 , \mu}^{(\lambda)}
        \right)               
       \Big| _{{\bm t}_{i} = {\bm t}_{i}^{*}}}
     {
       \prod_{\mu = 1}^{A}      
	   B \left( \alpha_{1 , \mu}^{(\lambda)} + \tilde{\alpha}_{1 , \mu}^{(\lambda)} ,
                \beta_{1 , \mu}^{(\lambda)} + \tilde{\beta}_{1 , \mu}^{(\lambda)}
        \right)
       B \left( \alpha_{0 , \mu}^{(\lambda)} + \tilde{\alpha}_{0 , \mu}^{(\lambda)} ,
                \beta_{0 , \mu}^{(\lambda)} + \tilde{\beta}_{0 , \mu}^{(\lambda)}
        \right)               
       \Big| _{{\bm t}_{i} = {\bm t}_{i}^{0}}}
\nonumber \\
& \ \ & \cdot
\frac{ \sum_{j = 1, j \neq i}^{N} t_{G_i^{0} j}^{0} }
     { \sum_{j = 1, j \neq i}^{N} t_{G_i^{*} j}^{0} }.
\end{eqnarray}

For the determination of the transition destination of $\bm \omega$,
it is not chosen probabilistically but always an inverted value of the current activity state,
because $\bm \omega$ is a binary variable taking 0 or 1.
Furthermore, since the value of $\bm \omega$ after the transition is also binary, no special change is required after transition. 
In contrast, $\bm \omega$ takes continuous value in our method, therefore it is necessary to change a decision rule 
for the transition destination.
In addition, for normalization, it is necessary to consider how to change the value of $\bm \omega$ after determining the transition destination.

Based on the fact that beta distribution is a special case of Dirichlet distribution, 
namely Dirichlet distribution of order two,
we design the decision rule for the transition destination of $\bm \omega$ and its value after the transition 
by considering the difference of update rule on $\bm t$ between BIM and ours.
Here we give the constraint that the number of dimensions does not increase, as for $\bm t$ 
in BIM in Eq (\ref{eq:update11sub}). As a result,
the transition destination is determined in proportion to the sum of the current activity values,
with the value of the concentration parameter being set to 0.
\begin{eqnarray}
\label{eq:TranRate_omega}
Q (G_{k \mu}^{*}| {\bm {\omega}_{\mu , \backslash k}^{0}})
& = &
\left\{
\begin{array}{ll}
\displaystyle \frac{\sum_{l = 1, l \neq k}^{M} {\omega}_{l \mu}^{0}}{M-1} 
& {\rm if} \ \ G_{k \mu}^* = 1, \vspace{2mm} \\
\displaystyle 
\frac{\sum_{l = 1, l \neq k}^{M} 1 - {\omega}_{l \mu}^{0}}{M-1}
& {\rm if} \ \ G_{k \mu}^* = 0.
\end{array}
\right.
\end{eqnarray}
For the value of $\bm \omega$ after the transition,
normalization is applied so that the sum of the activity values is 1.
Here the transition parameter $\alpha^{(w)}$ is introduced as in the case of $\bm t$
in Eq (\ref{eq:update11}), 
and the activity value of transition destination is increased by $\alpha^{(w)}$. 
\begin{eqnarray}
\label{eq:Tran_omega}
{\omega}_{k {\mu}}^{*}
& = &
\left\{
\begin{array}{ll}
\displaystyle \frac{{\omega}_{k \mu}^{0} + {\alpha}^{({\omega})}}
                   {1 + {\alpha}^{({\omega})}} & {\rm if} \ \ G_{k \mu}^* = 1, \vspace{2mm} \\
\displaystyle \frac{{\omega}_{k \mu}^{0} }
                   {1 + {\alpha}^{({\omega})}} & {\rm if} \ \ G_{k \mu}^* = 0.
\end{array}
\right.
\end{eqnarray}
Finally, acceptance rate is expressed by the formula in Eq (\ref{eq:AcceptRate_omega}).
\begin{eqnarray}
\label{eq:AcceptRate_omega}
a( {\omega}_{k \mu}^{*}, {\omega}_{k \mu}^{0})
& = &
\min \left(1 , 
\frac{P ( {\bm t}, {\bm \omega_{k \mu}^*}, {\bm \omega}_{\mu , \backslash k}^{0}, {\bm s} ) }
     {P ( {\bm t}, {\bm \omega}^{0}, {\bm s} ) } 
\frac{Q ( G_{k \mu}^{0} | {\bm \omega}_{\mu , \backslash k}^{0} ) }
     {Q ( G_{k \mu}^{*} | {\bm \omega}_{\mu , \backslash k}^{0} )}
     \right).
\end{eqnarray}

Once the variables are updated, the hyperparameters are also updated using updated ${\bm \omega}$ and ${\bm t}$. 
The hyperparameters to be updated are ${\alpha}_{\mu}^{(n)}$, which is related to membership of neuron, ${\alpha}_{\mu}^{(p)}, {\beta}_{\mu}^{(p)}$, which are related to ensemble activity, and ${\alpha}_{1 , \mu}^{(\lambda)}, {\beta}_{1, \mu}^{(\lambda)}, {\alpha}_{0 , \mu}^{(\lambda)}, {\beta}_{0, \mu}^{(\lambda)}$, which are related to synchronization between activities of ensemble and neurons. 
The update rules of the hyperparameters are given as follows,
\begin{eqnarray}
\label{eq:UpdateAlpha_n}
\hat{\alpha}_{\mu}^{(n)} 
 & = & {\alpha}_{\mu}^{(n)} + \tilde{\alpha}_{\mu}^{(n)}, 
\\
\label{eq:UpdateAlpha_p}
\hat{\alpha}_{\mu}^{(p)} 
 & = & {\alpha}_{\mu}^{(p)} + \tilde{\alpha}_{\mu}^{(p)},
\\
\label{eq:UpdateBeta_p}
\hat{\beta}_{\mu}^{(p)} 
 & = & {\beta}_{\mu}^{(p)} + \tilde{\beta}_{\mu}^{(p)},
\\
\label{eq:UpdateAlpha_lambda1}
\hat{\alpha}_{1, \mu}^{(\lambda)} 
 & = & {\alpha}_{1 , \mu}^{(\lambda)} + \tilde{\alpha}_{1 , \mu}^{(\lambda)},
\\
\label{eq:UpdateBeta_lambda1}
\hat{\beta}_{1, \mu}^{(\lambda)} 
 & = & {\beta}_{1 , \mu}^{(\lambda)} + \tilde{\beta}_{1 , \mu}^{(\lambda)},
\\
\label{eq:UpdateAlpha_lambda0}
\hat{\alpha}_{0, \mu}^{(\lambda)} 
 & = & {\alpha}_{0 , \mu}^{(\lambda)} + \tilde{\alpha}_{0 , \mu}^{(\lambda)},
\\
\label{eq:UpdateBeta_lambda0}
\hat{\beta}_{0, \mu}^{(\lambda)} 
 & = & {\beta}_{0 , \mu}^{(\lambda)} + \tilde{\beta}_{0 , \mu}^{(\lambda)},
\end{eqnarray}
where the symbols with tilde are defined in Eqs (\ref{eq:tildealpha})-(\ref{eq:tildebeta})
and the ones with hat are updated hyperparameters.

Finally, we mention the computational cost of our proposed method.
The computational cost of our method is larger than the original BIM by a factor of the number of clusters. 
However, in our analysis, the number of clusters is supposed to be relatively small. 
In addition, the computational cost of the original BIM is not so large. 
Therefore, the computational cost of our method is not a problem in practice.

\begin{algorithm}[t]               
\caption{MCMC algorithm}         
\label{alg:MCMC}                    
\begin{algorithmic}                  
\STATE{initialize $\bm t, \bm \omega$ according to uniform distribution}
\STATE{hyperparameters are set to the given initial values}
\FOR{execute until convergence of $\bm t, \bm \omega$}
	\FOR{each ensemble $\mu \in \{ 1, A \}$, $k \in \{ 1, M \}$ (in parallel if possible)} 
		\STATE{draw $G_{k \mu}^{*}$ by Eq \eqref{eq:TranRate_omega}}
		\STATE{calculate $\omega_{k \mu}^{*}$ by Eq \eqref{eq:Tran_omega}}
		\STATE{calculate acceptance rate by Eq \eqref{eq:AcceptRate_omega}}
	\ENDFOR
	\FOR{each ensemble $\mu \in \{ 1, A \}$, $k \in \{ 1, M \}$ (updating synchronously)}
		\STATE{Update the value of $\bm \omega$ synchronously according to the computed acceptance rate}
	\ENDFOR
	\FOR{each neuron $i \in \{ 1, N \}$ (in parallel if possible)}
		\STATE{draw $G_{i}^{*}$ by Eq \eqref{eq:TranRate_t}}
		\FOR{each ensemble $\mu \in \{ 1, A \}$}
			\STATE{calculate ${\bm t}_{\mu}^{*}$ by Eq \eqref{eq:Tran_t}}
		\ENDFOR
		\STATE{calculate acceptance rate by Eq \eqref{eq:AcceptRate_t}}
	\ENDFOR
	\FOR{each neuron $i \in \{ 1, N \}$ (updating synchronously)}	
		\STATE{update the values of $\{ t_{1i}, \ldots, t_{Ai} \}$ synchronously according to the computed acceptance rate}
	\ENDFOR
		\STATE{update hyperparameter values using updated $\bm \omega$ and $\bm t$ by following Eqs \eqref{eq:UpdateAlpha_n}-\eqref{eq:UpdateBeta_lambda0}}
\ENDFOR
\end{algorithmic}
\end{algorithm}

\section*{Results and Discussion}
\label{chap:result}

\subsection*{The model of fluorescence intensity}

We conduct numerical analysis for the effectiveness and the validity of our proposed method by applying to continuous-valued data.
The synthetic activity data for validation of the proposed method
is generated under consideration of experimental fluorescence intensity 
by calcium imaging.
In calcium imaging, fluorescent protein binds to calcium ions and emits light, and its activity is observed as light intensity. 
There is a chemical nonlinear relation between neuronal activity and fluorescence intensity.
This relation depends on calcium ion concentration and the type of fluorescent protein, 
and nonlinearity is controlled by the parameters in general.

In our study, leaky integrated-and-fire model is used to generate the synthetic data.
This mathematical model is for representing time series data of electrical intracellular membrane potentials. 
In this model, the relation among inflow currents from other neurons via synaptic and gap junctions,
the external stimulus currents, and the time variation of membrane potential is 
expressed as differential equations with additional noise term, 
\begin{eqnarray}
\left\{
\begin{array}{ll}
\label{eq:Leaky}
\tau_{i}^{(m)} \frac{\displaystyle dV_{i \tau}}{\displaystyle d \tau}
= 
- (V_{i \tau} - V_{i}^{({\rm rest})}) & \\
\hspace{2cm} +  R_{i} ( I_{i \tau}^{({\rm chem})}  + I_{i \tau}^{({\rm gap})} + I_{i \tau}^{({\rm stim})} )
+ {\sigma}_{i}^{({\rm noise})} {\xi}_{i} (\tau)
& {\rm if}  \ \ V_{i \tau} < V_{i}^{({\rm th})}, \\
V_{i \tau}  =  V_{i}^{({\rm act})}, V_{i (\tau+d\tau)} = V_{i}^{({\rm init})} 
& {\rm if} \ \ V_{i \tau} \geq V_{i}^{({\rm th})},
\end{array}
\right.
\end{eqnarray}
where the definitions of variables and parameters are given in Table \ref{tab:Leaky}.

The input currents via chemical synaptic junction and gap junction are determined by
the chemical synaptic junction weight $w_{ij}^{\rm (chem)}$ and the gap junction weight $w_{ij}^{\rm (gap)}$, respectively.
The current via chemical synaptic junction is given by the logistic function of membrane potential in Eq (\ref{eq:currentchem}). 
For the current via the gap junction, the weight $w_{ij}^{\rm (gap)}$ serves as a constant resistance in Eq (\ref{eq:currentgap}).
To summarize, these currents are expressed as
\begin{eqnarray}
I_{i \tau}^{({\rm chem})}
& = &
\sum_{j = 1}^{N} 
\frac{w_{i j}^{({\rm chem})}}
     {1 + {\exp} \left[ - \frac{(V_{j \tau} - V^{({\rm half})})}{V^{({\rm width})}}  \right] },
\label{eq:currentchem}
\\
I_{i \tau}^{({\rm gap})}
& = &
\sum_{j = 1}^{N} w_{i j}^{({\rm gap})} (V_{j \tau} - V_{i \tau} ),      
\label{eq:currentgap}
\end{eqnarray}
where $V^{(\rm half)}$ and $V^{(\rm width)}$ represent the position of half value
and half width of the logistic function, respectively.
For chemical synaptic connection, the sign of $w_{i j}^{\rm (chem)}$ is positive for excitatory connection and negative for inhibitory, 
respectively.
In numerical analysis, these equations are implemented by Euler-Maruyama method to generate time series signals 
of membrane potentials.

\begin{table}[t]
\begin{center}
  \caption {Variables and parameters in leaky integrated-and-fire model}
  \label{tab:Leaky}
  \begin {tabular}{c||l} \hline 
 $V_{i \tau}$ & membrane potential of the $i$th neuron at time $\tau$ \\
 $\tau_i^{(m)}$ & time constant of the $i$th neuron \\
 $V_i^{\rm (rest)}$ & static membrane potential for the $i$th neuron \\
 $\sigma_i^{\rm (noise)}$ & magnitude of noise for the $i$th neuron \\
 $\xi_i (\tau)$ & noise term of the $i$th neuron (white standard Gaussian noise) \\
 $R_i$ & membrane resistance of the $i$th neuron \\
 $V_i^{\rm (th)}$ & firing threshold for the $i$th neuron \\
 $V_i^{\rm (act)}$ & action potential of the $i$th neuron \\
 $V_i^{\rm (init)}$ & threshold of membrane potential for the $i$th neuron \\
 $\tau$ & time \\
 $d\tau$ & time interval \\
 $I_{i \tau}^{\rm (chem)}$ & input current through chemical synaptic junction to the $i$th neuron at time $\tau$ \\
 $I_{i \tau}^{\rm (gap)}$ & input current through gap junction to the $i$th neuron at time $\tau$ \\
 $I_{i \tau}^{\rm (stim)}$ & input current by external stimulus to the $i$th neuron at time $\tau$ \\
 \hline
  \end {tabular}
\end{center}
\end{table}

Next, the model for converting membrane potential to fluorescence 
intensity is described.
The time derivative of fluorescence intensity in the calcium imaging is given 
by the following differential equation,
\begin{eqnarray}
 \tau_i^{(F)} \frac{dF_{i\tau}}{d\tau} = f ([{\rm Ca}^{2+}]_{i \tau}) - F_{i\tau},
 \label{eq:difffluo}
\end{eqnarray}
where $\tau_i^{(F)}$ is the time constant of the fluorescent protein, $[{\rm Ca}^{2+}]_{i \tau}$ is the intracellular calcium ion 
concentration of the $i$th neuron at time $\tau$, and $F_{i \tau}$ is the fluorescence intensity of the $i$th neuron at time $\tau$.
The nonlinear function for calcium ion concentration in Eq (\ref{eq:difffluo}) is expressed by sigmoid-like function, which is called Hill's equation as
\begin{eqnarray}
f ([{\rm Ca}^{2+}]_{i \tau}) 
=
\left\{
\begin{array}{ll}
F^{(\rm max)} 
\frac{ ( [{\displaystyle \rm Ca}^{2+}]_{i \tau} )^h }{\displaystyle ( K^{(D)})^{h} + ([{\rm Ca}^{2+}]_{i \tau})^h }
& {\rm if} \ \ [{\rm Ca}^{2+}]_{i \tau} \geq 0, \\
0
& {\rm if}  \ \ [{\rm Ca}^{2+}]_{i \tau} < 0,
\end{array}
\label{eq:Hill}
\right.
\end{eqnarray}
where $F^{(\rm max)}$ is the maximum of fluorescence intensity, $K^{(D)}$ and $h$ are parameters for controlling the shape of
sigmoid-like function.
The chemical response properties of fluorescent proteins to changes in calcium concentration 
are described in Ref \cite{Ca2indicator}. We used the function in Eq (\ref{eq:Hill}) by referring it.

In generating the data of fluorescence intensity, the relation between calcium ion concentration and membrane potential is
 simplified in our work. Namely, we assume that calcium ion concentration is directly proportional to 
 membrane potential as $[{\rm Ca}^{2+}]_{i \tau} \propto V_{i \tau}$ (we set the constant of proportionality unity in the following), 
 which leads to the differential equation for fluorescence intensity in the following form,
\begin{eqnarray}
\tau_{i}^{(F)} \frac{dF_{i \tau}}{d\tau}
= f(V_{i \tau}) - F_{i \tau}.
\end{eqnarray}
In applying our proposed method to synthetic fluorescence intensity practically, 
the fluorescence intensity data is rescaled to $\tilde{F}_{i \tau}$ as follows,
 whose value is in the range $[0,1]$.
\begin{eqnarray}
\tilde{F}_{i \tau} 
= 
\frac{F_{i \tau} - \min_{1 \leq \tau \leq M} F_{i \tau}}
     {\max_{1 \leq \tau \leq M} F_{i \tau} - \min_{1 \leq \tau \leq M} F_{i \tau}}.
\end{eqnarray}

\subsection*{Generation of synthetic data}

\begin{table}[t]
\begin{center}
\caption{Values of parameters in leaky integrated-and-fire model: The unit of each quantity can be
chosen auxiliary in this numerical analysis. In practical application, the unit can be chosen
appropriately to match the experimental condition.}

\label{tab:ParaDef}
\begin{tabular}{c|c||c|c}  
  parameter
& value 
& parameter
& value \\ \hline \hline
  $\tau_{i}^{(m)}$ (${\rm E}_1 \sim {\rm E}_4$) 
& $0.05$ 
&  $w_{ij}^{({\rm chem})}$ (excitatory)
& $0.08$ \\
  $\tau_{i}^{(m)}$ (${\rm E}_5 \sim {\rm E}_8$) 
& $0.1$ 
& $w_{ij}^{({\rm chem})}$ (inhibitory)
& $-0.08$ \\
  $V_{i}^{({\rm rest})}$
& $0$
& $w_{ij}^{({\rm gap})}$
& $0.3$ \\
  ${\sigma}_{i}^{({\rm noise})}$
& $0.2$ 
& $V^{({\rm half})}$ 
& $0.3$ \\
  $R_{i}$
& $1$  
& $V^{({\rm width})}$ 
& $0.1$ \\
  $V_{i}^{({\rm th})}$
& $0.2$ 
& $F^{({\rm max})}$ 
& $5$ \\
  $V_{i}^{({\rm act})}$
& $1$ 
& $K^{({\rm D})}$ 
& $0.3$  \\
  $V_{i}^{({\rm init})}$
& $0$ 
& $h$ 
& $3$ \\
  $I_{i}^{({\rm stim})}$
& $0.3$ 
& $\tau_i^{(F)}$ 
& $0.1$ \\
\hline
\end{tabular}
\end{center}
\end{table}

\begin{table}
\begin{center}
\caption{Probability of connections between structural ensembles}
\label{tab:ComDef}
\begin{tabular}{c||c}  
type of connection between neurons & probability \\ \hline \hline
  synaptic connections between neurons in the same ensemble 
& $0.5$  \\
  excitatory synaptic connection across ensembles: ${\rm E}_1 \rightarrow {\rm E}_2$  
& $0.5$  \\
  inhibitory synaptic connection across ensembles: ${\rm E}_1 \rightarrow {\rm E}_3$  
& $0.5$  \\
  excitatory synaptic connection across ensembles: ${\rm E}_6 \rightarrow {\rm E}_5$
& $0.5$  \\
  inhibitory synaptic connection across ensembles: ${\rm E}_7 \rightarrow {\rm E}_5$  
& $0.5$  \\
  gap junction between neurons in the same ensemble  
& $0.8$  \\
\hline
\end{tabular}
\end{center}
\end{table}

In our numerical analysis, we generate synthetic data with $N=100, M=2000, A=8$, and with "structural" ensembles ${\rm E}_1 \sim {\rm E}_8$ 
defined by the geometry of structural connections.
Two structural ensembles ${\rm E}_1$ and ${\rm E}_5$ have 20 neurons and others have 10 neurons.
 
\begin{figure}[t]
\begin{picture}(100,265)
\put(0,10){\includegraphics[scale=0.3]{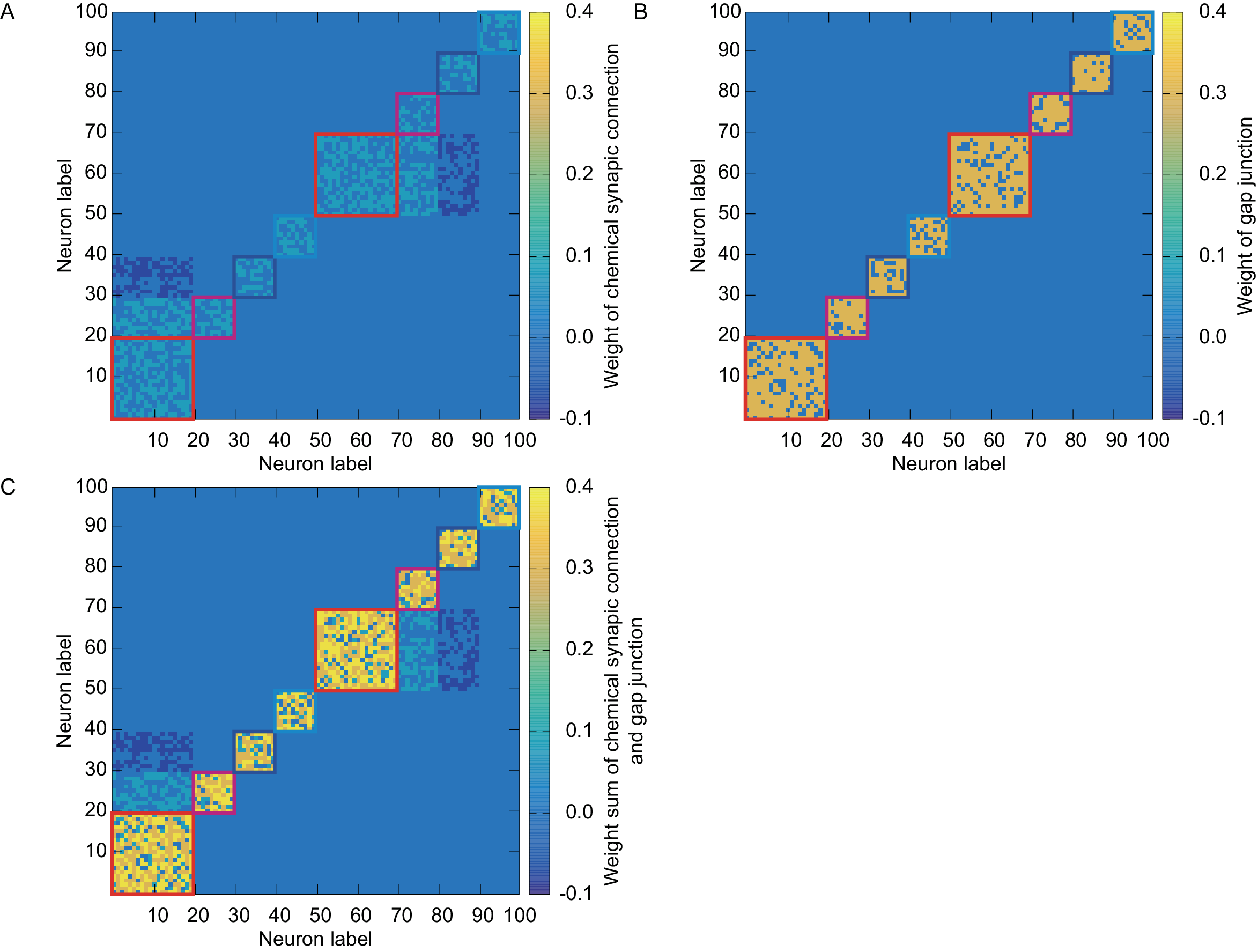}}
\end{picture}
\caption{The connection matrices between neurons: 
A: Weight of chemical synaptic connection $w_{ij}^{({\rm chem})}$.
B: Weight of gap junction $w_{ij}^{({\rm gap})}$.
C: Weight sum of chemical synaptic connection and gap junction $w_{ij}^{({\rm chem})}+w_{ij}^{({\rm gap})}$.}
\label{fig:WAll}
\end{figure}

The values of parameters in leaky integrated-and-fire model are common to all neurons as in Table \ref{tab:ParaDef}.
For fluorescent protein, we set parameters for chemical reaction by considering specific fluorescent protein.
Neurons in the structural ensembles ${\rm E}_1 \sim {\rm E}_4$ respond faster than the ones in ${\rm E}_5 \sim {\rm E}_8$ 
because their time constant $\tau_i^{(m)}$ is smaller. Between ensembles,
structural connections are given probabilistically as synaptic connection and gap junction.
Excitatory synaptic connections are given in the ensemble pairs, ${\rm E}_1 \rightarrow {\rm E}_2$ and ${\rm E}_6 \rightarrow {\rm E}_5$. 
Similarly, inhibitory synaptic connections are given in the pairs ${\rm E}_1 \rightarrow {\rm E}_3$ and ${\rm E}_7 \rightarrow {\rm E}_5$.
There is no synaptic connection between ensembles except for the above-mentioned pairs.
In addition, excitatory synaptic connections and symmetric gap junctions are given probabilistically between neurons in the same ensemble.
The connection probabilities are given in Table \ref{tab:ComDef}. 
As a result, we obtain the connection matrices of chemical synaptic connection and gap junction between neurons as shown in Fig \ref{fig:WAll}.

In out setting, the reason for introducing disconnected structural ensembles ${\rm E}_4$ and ${\rm E}_8$ is as follows. 
As mentioned in the introduction, 
our method estimates functional neuronal ensembles only from synchronization of activity data.
Therefore, it cannot directly find the possible background structural connection between neurons. 
Even though the result of estimation for functional ensembles reflects the geometry of structural ensembles,
functional and structural ensembles are not always the same.
To demonstrate the difference between functional and structural ensembles, 
we intentionally use the same input currents to obtain similar neuronal activities 
for prepared structural ensembles, where some ensembles have inter-ensemble connections and others do not.

In addition, due to the difference between functional and structural ensembles as mentioned above, the exact structural ensembles cannot be found by our method. Hence, it is not necessary to introduce detailed structure of neuronal network geometry in the numerical analysis.
For investigating the basic properties of the proposed method, 
we use simple neuronal network geometry with 8 structural ensembles in our numerical analysis. 

\begin{figure}[H]
\begin{picture}(100,220)
\put(0,10){\includegraphics[scale=0.16]{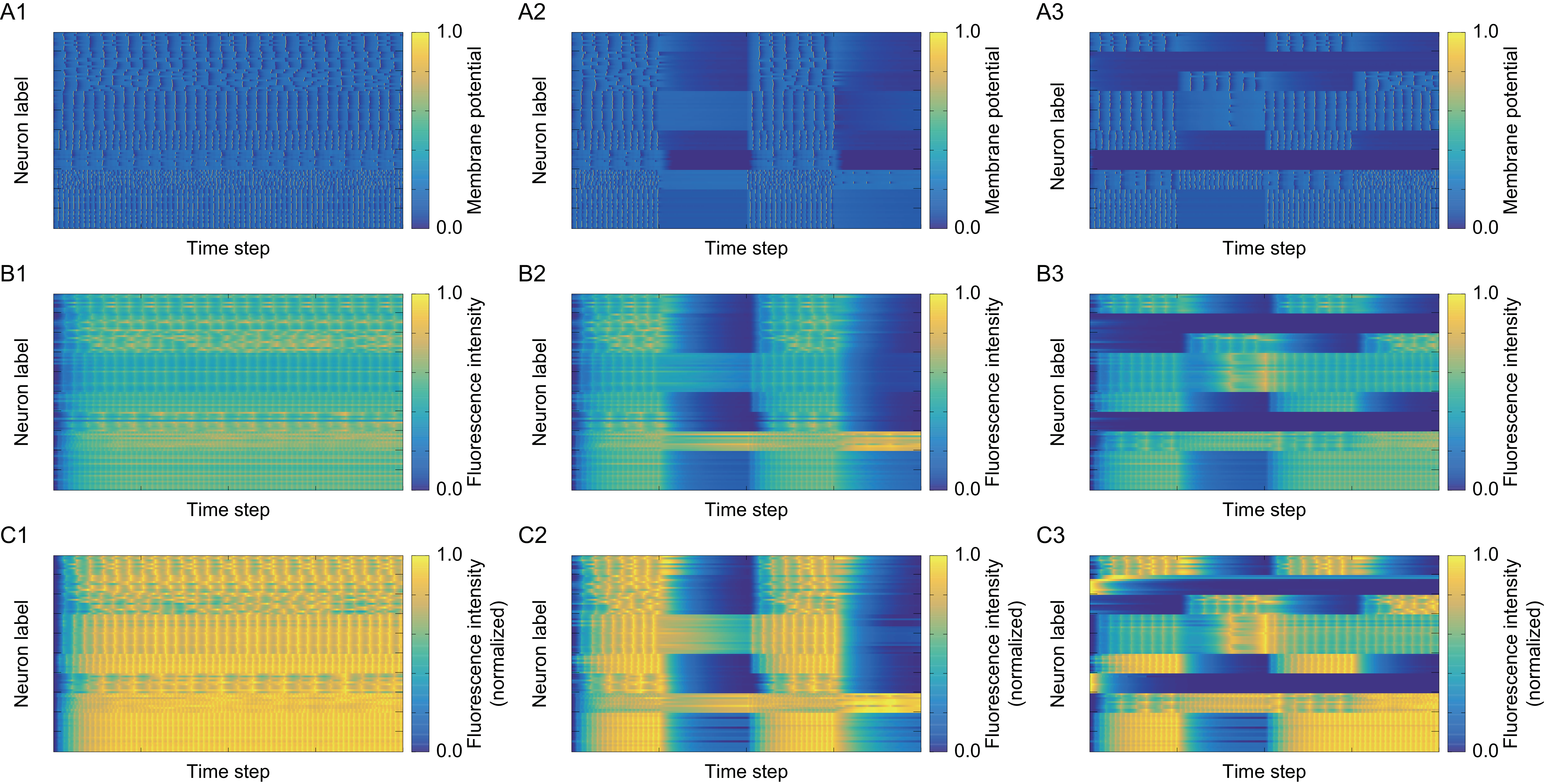}}
\end{picture}
\caption{ 
The signals in leaky integrated-and-fire model: In the identifier of each figure, the alphabet and the number
represent the information of the signal and the input, respectively. The details are as follows.
A: Membrane potential $V_{i\tau}$.
B: Fluorescence intensity $F_{i\tau}$.
C: Normalized fluorescence intensity $\tilde{F}_{i\tau}$.
1: Identical and stationary input.
2: Identical and non-stationary input.
3: Different and non-stationary input.}
\label{fig:NeuroActAll}
\end{figure}

\begin{figure}[H]
\begin{picture}(100,250)
\put(0,10){\includegraphics[scale=0.5]{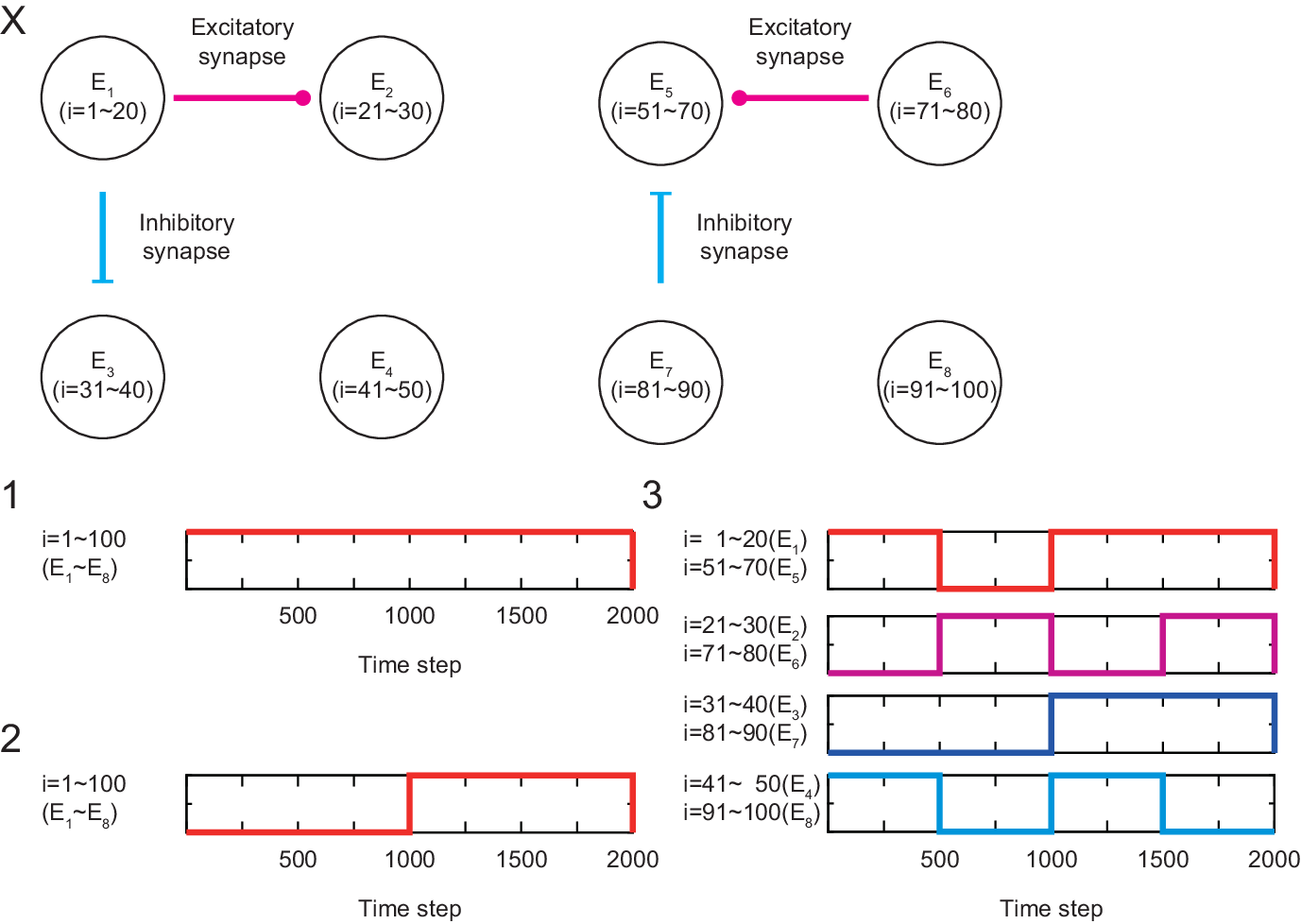}}
\end{picture}
\caption{ 
The ensembles in neuronal network and the input signals in the numerical analysis:
X: Structural ensembles.
1: Identical and stationary input current.
2: Identical and non-stationary input current.
3: Different and non-stationary input current.}
\label{fig:EnsStrAndInput}
\end{figure}

For investigating stationary and non-stationary activities, we generate activity data with constant and time-varying input currents. Specifically, we generate three time series data, which are shown in Fig \ref{fig:NeuroActAll}: the data under the same and stationary input current to all neurons as in 1 in Fig \ref{fig:EnsStrAndInput}, 
the data under the same and non-stationary input current to all neurons as in 2 in Fig \ref{fig:EnsStrAndInput}, 
and the data under different non-stationary input current to each ensemble as in 3 in Fig \ref{fig:EnsStrAndInput}.
The non-stationary input currents are the same in the structural ensembles pairs of $\{ {\rm E}_1, {\rm E}_5 \}$, 
$\{ {\rm E}_2, {\rm E}_6 \}$, $\{ {\rm E}_3, {\rm E}_7 \}$, and $\{ {\rm E}_4, {\rm E}_8 \}$.
In Fig \ref{fig:NeuroActAll}, membrane potential $V_{i\tau}$ under these conditions is shown in A, fluorescence intensity 
$F_{i\tau}$ in B, 
and rescaled fluorescence intensity $\tilde{F}_{i\tau}$ in C.
As seen in Fig \ref{fig:NeuroActAll}, the activities are similar for neurons in the ensemble pair with the same pattern of input currents.

\begin{figure}[H]
\begin{picture}(100,150)
\put(0,10){\includegraphics[scale=0.16]{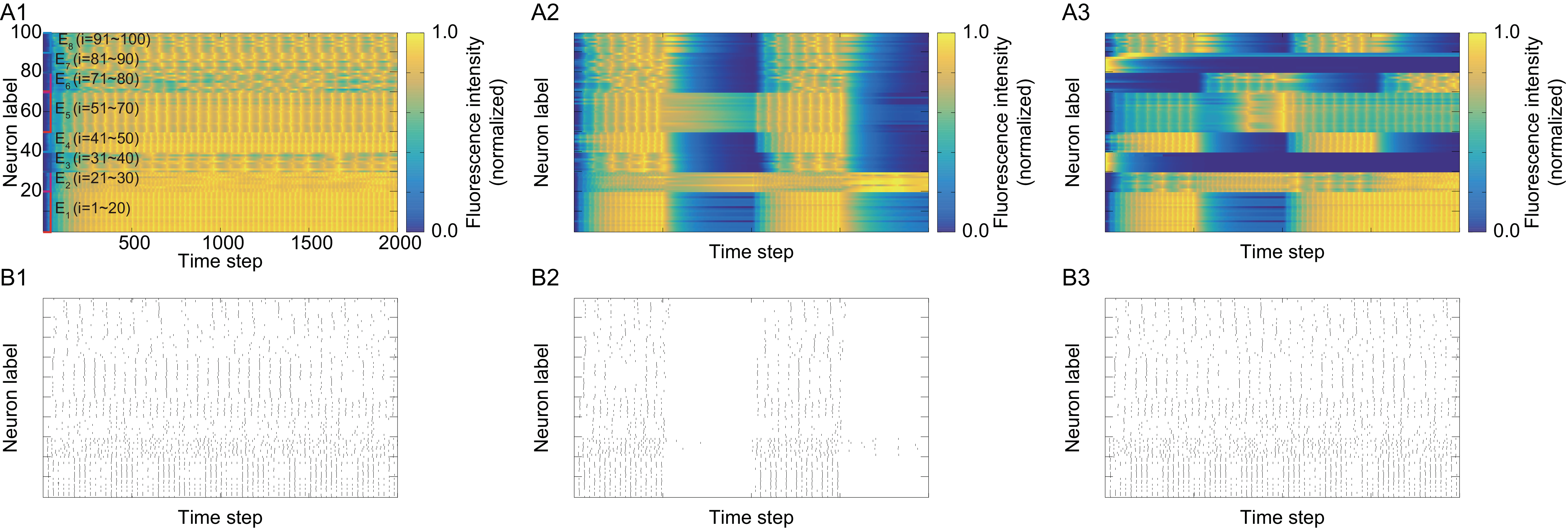}}
\end{picture}
\caption{Synthetic activity data to be analyzed:
The figures A1, A2, and A3 describing normalized fluorescence intensity $\tilde{F}_{i\tau}$  
are the same as C1, C2, and C3 in Fig \ref{fig:NeuroActAll}, respectively.
The conditions in leaky integrated-and-fire model are summarized as follows.
A1: Stationary activity data under the same stationary input current to all neurons.
A2: Non-stationary activity data under the same non-stationary input current to all neurons.
A3: Non-stationary activity data under different non-stationary input currents to each ensemble.
The figures B1, B2, B3 are binarized data of A1, A2, A3, respectively.}
\label{fig:Act}
\end{figure}

\begin{table}[H]
\begin{center}
  \caption {Conditions of numerical inference by MCMC}
  \label {tab:Condition}
  \begin {tabular}{c||c|c} \hline
     parameter for inference & BIM & our proposed method \\ \hline \hline
    number of initial ensembles: $A_{\rm init}$ & 50 & $16$  \\
    number of trials: $r_{\rm max}$ & $10$ & $10$ \\
    number of updates in MCMC: $ {\gamma}_{\rm max}$  & $1000$ & $1000$ \\
    transition parameter: $\alpha^{(t)}$ & none & $\{ 0.1, 0.3, 0.5, 0.8 \}$  \\
    transition parameter: $\alpha^{(w)}$ & none & $1000$  \\ \hline
  \end {tabular}
\end{center}
\end{table}

We should compare the result of our numerical analysis with BIM, however application of the inference method in BIM is limited to binary data. 
Therefore, when applying two methods to synthetic continuous-valued data, raw continuous data is used in our proposed method and the binarized one is used in BIM.
Binarization is performed using Peakfind function in MATLAB, where local maximum of time-series signal is regarded as spike.
More precisely, Peakfind function can find local maximum value of a sample point in a data series, which is larger 
than the values of two adjacent samples and considered as a local peak. 
For the detail of Peakfind function, see the reference of MATLAB. 

The figures A1, A2, and A3 in Fig \ref{fig:Act} show synthetic activity data under the parameters
in Tables \ref{tab:ParaDef} and \ref{tab:ComDef}.
In contrast, B1, B2, and B3 are the data after binarization using Peakfind function, which are for the numerical analysis of BIM.
In B1, B2, and B3, the activities in the same structural ensemble are also similar as in A1, A2, and A3. Therefore we consider there is no problem with data preprocessing by binarization.
Furthermore, it can be observed that the activities of neurons in the downstream structural ensembles, namely ${\rm E}_2, {\rm E}_3, {\rm E}_5$ in our setting, tend to be disturbed by the neurons in the upstream ensembles, ${\rm E}_1, {\rm E}_6, {\rm E}_7$.
More precisely, synchronization of neuronal activities in the downstream ensemble tends to be disrupted due to the activity of neurons in the upstream ensemble, as can be seen in A1, A2, and A3 in Fig \ref{fig:Act}.
 
\subsection*{Results of the application and discussion}

As mentioned before, the purpose of our method is to infer functional neuronal ensembles, 
and it does not matter whether the structural neuronal ensembles defined by structural connections can be estimated or not. 
In other words, there is no ground-truth solution. 
Therefore, we attempt to capture neuronal ensembles with similar activities under various "granularities", 
namely scales of ensembles.
We emphasize that our soft clustering approach allows ensemble inference at different granularities, 
which is not possible in BIM due to hard clustering. 
Since our goal is to estimate the ensembles at different granularities, we cannot judge the superiority or inferiority of ensemble inference performance using a metric such as the distance from true structural connection. Hence, we display the results visually and discuss them. 
Additionally, stationary and non-stationary data are used to show that even non-stationary data can be handled by our soft clustering
method. 

We apply two methods to synthetic data: our proposed method and BIM.
For the performance of functional ensemble inference, the average of multiple trials with different initial conditions is taken, because the dependence of MCMC result on initial condition should be removed.
In BIM, the number of initial ensembles is set to half the number of neurons as recommended in \cite{diana2019bayesian}, which is to avoid convergence to an inappropriate local solution.
Other conditions are the same in two methods.
The detailed conditions of the numerical analysis are summarized in Table \ref{tab:Condition}.

For the behavior by multiple trials, we observe how often each neuron is classified into the same functional ensemble, 
for which we define the similarity matrix ${\bm U}$. The element of $\bm U$ represents the similarity of activities 
between neurons, which has the information of functional ensembles averaged over multiple trials.
 Element of the similarity matrix is expressed as $U_{ij} = \sum_{r=1}^{r_{\rm max}} \sum_{\mu =1}^{A_{\rm init}} t_{\mu i}^{r} t_{\mu j}^{r}$, where superscript $r$ is the label of trial number with initial condition being changed.

\begin{figure}[H]
\begin{picture}(100,370)
\put(0,10){\includegraphics[scale=0.20]{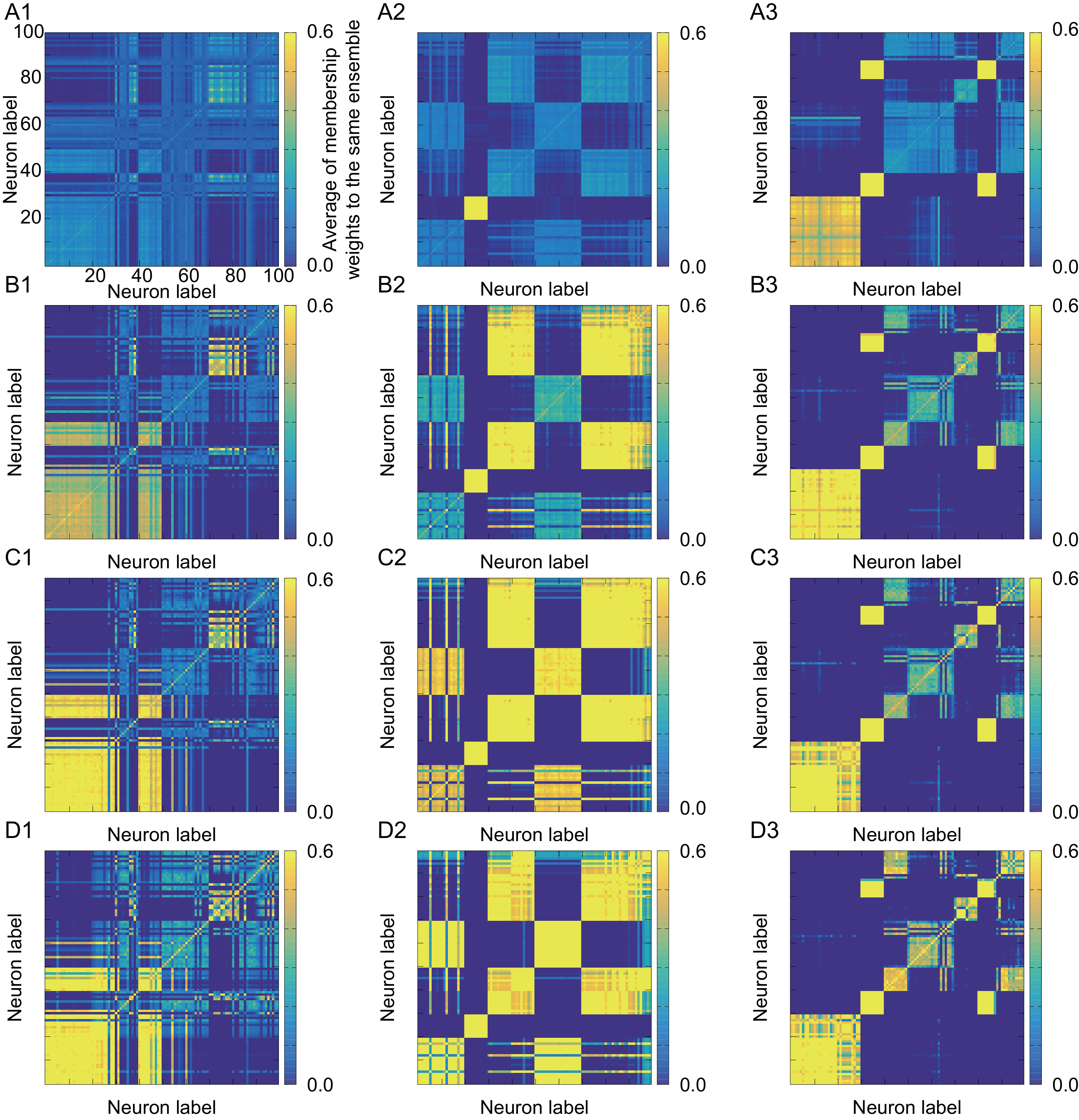}}
\end{picture}
\caption{Heat map of $\bm U$ by applying our proposed method to stationary and non-stationary activities:
In each figure, the alphabet and the number in the identifier
represent the information of transition parameter $\alpha^{(t)}$ and the input signal, respectively. The details are as follows.
A: $\alpha^{(t)} = 0.1$.
B: $\alpha^{(t)} = 0.3$.
C: $\alpha^{(t)} = 0.5$.
D: $\alpha^{(t)} = 0.8$.
1: Identical and stationary input.
2: Identical and non-stationary input.
3: Different and non-stationary input.}
\label{fig:ResultProp}
\end{figure}

By comparing A3, B3, C3, and D3 in Fig \ref{fig:ResultProp} under
different and non-stationary input current as in 3 in Fig \ref{fig:EnsStrAndInput}, 
the resulting functional ensembles are changed by the value of the transition parameter $\alpha^{(t)}$. 
For example, when focusing on neurons $i=1 \sim 30$ in the structural ensembles ${\rm E}_1$ and ${\rm E}_2$, 
the coarse-grained large functional ensemble is obtained under small $\alpha^{(t)}$ as in A3 in Fig \ref{fig:ResultProp}, where the original two structural ensembles ${\rm E}_1$ and ${\rm E}_2$ are merged into one functional ensemble. 
In contrast, smaller functional ensembles are obtained under large $\alpha^{(t)}$ as in D3 
in Fig \ref{fig:ResultProp}, where the structural ensembles ${\rm E}_1$ and ${\rm E}_2$ are separated.
Furthermore, in our method, we observe that the neurons in different structural ensembles, 
for example the neurons in the structural ensemble pair $\{ {\rm E}_3, {\rm E}_7 \}$ or another pair $\{ {\rm E}_4, {\rm E}_8 \}$, are classified into the merged functional ensemble in all cases of $\alpha^{(t)}$, although there is no structural connection between ensembles
in the pair. 
The neurons in such ensemble pair have similar neuronal activity as in A3 in Fig \ref{fig:Act}.
This is because the same input current is provided into the ensemble pair as in 3 in Fig \ref{fig:EnsStrAndInput}.
In our method, the inference of functional ensemble is based only on synchronization of activity. 
Thus, even if there is no structural connection, these neurons with similar activity 
are classified into the same functional ensemble as a consequence. 

\begin{figure}[H]
\begin{picture}(100,110)
\put(0,10){\includegraphics[scale=0.20]{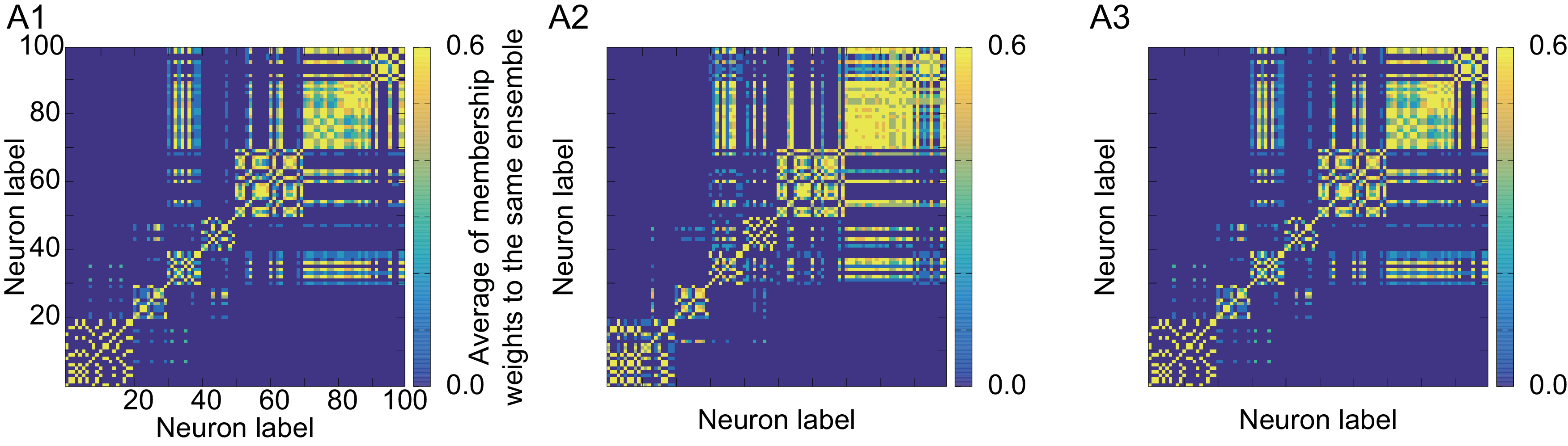}}
\end{picture}
\caption{Heat map of $\bm U$ by applying BIM to stationary and non-stationary activities: 
A1: Identical and stationary input.
A2: Identical and non-stationary input.
A3: Different and non-stationary input.}
\label{fig:ResultPrev}
\end{figure}

In contrast, when the method in BIM is applied to the activity data after binarization, 
only the course-grained functional ensembles can be found as in A3 in Fig \ref{fig:ResultPrev}. 
It is difficult to separate the original structural ensembles with similar activity by this method, because this method can only be used as hard clustering and does not have an extra parameter to control hard/soft clustering like the transition parameter in our proposed method.

We also observe the dependence on the input current.
In our proposed method, when the activities in all neurons are similar as in A1 in Fig \ref{fig:Act}, 
the difference in time constants $\tau_i^{(m)}$ between the ensemble groups, $\{ {\rm E}_1, {\rm E}_2, {\rm E}_3, {\rm E}_4 \}$ 
and $ \{ {\rm E}_5, {\rm E}_6, {\rm E}_7, {\rm E}_8 \}$, is 
mainly reflected in resulting course-grained functional ensembles. 
The differences among ensembles in the same group can be found in more detail as we switch toward hard clustering
(or large $\alpha^{(t)}$). This behavior can be observed in A1, B1, C1, and D1 in Fig \ref{fig:ResultProp}.
In contrast, in the case of A2 in Fig \ref{fig:ResultProp}, 
difference in information flow between ensembles due to non-stationary input current leads to clear difference in activity. 
In this case, a large functional ensemble is confirmed as the merged structural ensemble groups $ \{ {\rm E}_3, {\rm E}_4, {\rm E}_6, {\rm E_7}, {\rm E}_8 \}$, which have similar activities. 
As we switch closer to hard clustering, the hierarchy can be found: the smaller functional ensembles of $\{ {\rm E}_3$, ${\rm E}_6 \}$ 
and $ \{ {\rm E}_4, {\rm E}_7, {\rm E}_8 \}$
(see D2 in Fig \ref{fig:ResultProp}).
On the other hand, the structural ensembles in the groups of $\{ {\rm E}_3, {\rm E}_6 \}$ or 
$\{ {\rm E}_4, {\rm E}_7, {\rm E}_8 \}$ cannot be separated anymore, where the activities of neurons highly synchronize in each group.

While BIM can estimate course-grained functional ensemble,
it cannot find the hierarchy of ensembles 
confirmed by our method from the results in A1, A2, and A3 in Fig \ref{fig:ResultPrev}.
Another feature is that the presence or absence of stationarity has little influence on the results by BIM.
This may be due to hard clustering or ensemble inference by time-averaged activity, because
hard clustering or time-averaged activity cannot appropriately capture the temporal change of the feature in the activity. 

Both of BIM and our proposed method determine
functional ensembles based on synchronization of activity. Therefore, 
the activation timings must coincide among neurons to be regarded as synchronous activity in both models.
Nevertheless, our method has the advantage for functional ensemble inference.
There are upstream and downstream information transmissions  
through connections. The downstream neurons are activated 
later than the upstream ones, and there is always a time gap even in the activity of synchronous neurons.
In binary data, the perfect coincidence of activation timing (namely spike timing) is required for identifying synchronization. 
Hence, it is difficult to find synchronization due to the presence of time gap.
In contrast, in continuous-valued data, the activity data before and after the maximum activity value (or spike) can also be 
used to identify synchronization.
Therefore, the model of continuous activity can identify synchronization more appropriately than the binary model.
We guess that this is the reason that the functional ensemble in Fig \ref{fig:ResultPrev} by BIM
becomes always similar regardless of the feature of external input current.

\begin{figure}[H]
\begin{picture}(100,500)
\put(0,10){\includegraphics[scale=0.70]{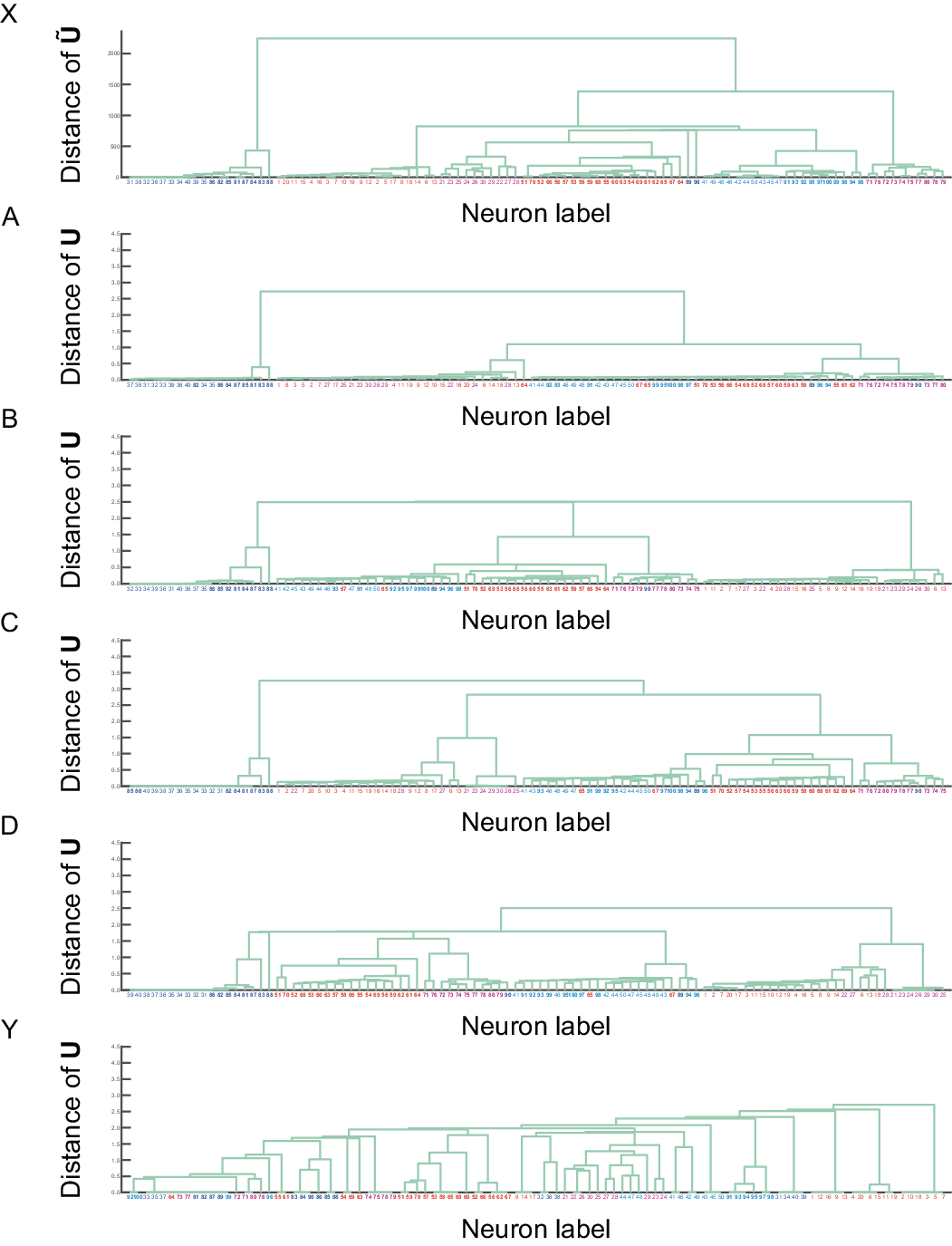}}
\end{picture}
\caption{Dendrogram of hierarchical structure for functional ensembles:
Each dendrogram is depicted using different similarity matrix for computing distance between nodes.
X: Matrix $\tilde{\bm U}$ from activity data in A3 in Fig \ref{fig:Act}.
A: Matrix $\bm U$ in A3 ($\alpha^{(t)}=0.1$) in Fig \ref{fig:ResultProp}.
B: Matrix $\bm U$ in B3 ($\alpha^{(t)}=0.3$) in Fig \ref{fig:ResultProp}.
C: Matrix $\bm U$ in C3 ($\alpha^{(t)}=0.5$) in Fig \ref{fig:ResultProp}.
D: Matrix $\bm U$ in D3 ($\alpha^{(t)}=0.8$) in Fig \ref{fig:ResultProp}.
Y: Matrix $\bm U$ in A3 (BIM) in Fig \ref{fig:ResultPrev}. 
The vertical and horizontal axes indicate distance and neuron label, respectively.
The color of neuron label indicates the color of input current shown in 3 in Fig \ref{fig:EnsStrAndInput}.
For example, neuron labels in the structural ensembles $\{ {\rm E_1}, {\rm E_5} \}$ with the same input current 
are shown by the same color.
The font of neuron label in $\{ {\rm E_1}, {\rm E_2}, {\rm E_3}, {\rm E_4} \}$ (ensembles with smaller $\tau_i^{(m)}$)
 is shown by lightface, while in $\{ { \rm E_5 }, {\rm E_6}, {\rm E_7}, {\rm E_8} \}$ (with larger $\tau_i^{(m)}$) 
 by boldface.
}
\label{fig:dendrogram}
\end{figure}

Finally, for comparison between the results by our proposed method and BIM,
dendrograms for hierarchical structure of functional ensembles are depicted in Fig \ref{fig:dendrogram},
where the data under non-stationary input current (3 in Fig \ref{fig:EnsStrAndInput}) is used. 
For dendrograms, similarity matrices $\bm U$ in Figs \ref{fig:ResultProp} and \ref{fig:ResultPrev} 
are used for computing distance between neurons.
In addition, we also define similarity matrix $\tilde{\bm U}$ for input signal, namely neuronal activity $\bm s$ as
\begin{equation}
\tilde{U}_{ij} = \sum_{k=1}^{M} s_{ik} s_{jk}.
\end{equation}

From dendrogram X in Fig. \ref{fig:dendrogram} depicted by $\tilde{\bm U}$ from activity data in A3 in Fig \ref{fig:Act},
it is found that the activities of neurons in structural ensembles under the same input current are similar, 
even if the structural ensembles are not directly connected.
Next, from dendrograms A, B, C, and D depicted by $\bm U$ under different $\alpha^{(t)}$,
hierarchical structure of dendrogram is changed by tuning
transition parameter $\alpha^{(t)}$. Obviously, the structures of dendrograms A, B, C, and D are similar to 
X. This means that the hierarchical structure of function ensembles by our proposed method 
reflects the hierarchical structure of input activity data correctly, which validates the result by our proposed method. 
For small $\alpha^{(t)}$ or soft clustering case,
the distance between neurons tends to be small. 
This implies that membership weight $\bm t$ of individual neuron spreads over many ensembles,
and the number of neurons in one ensemble tends to be large.
In contrast, for large $\alpha^{(t)}$ or hard clustering case, distance between neurons tends to be large.
In this case, the membership weight $\bm t$ concentrates on a specific ensemble, and the number of
neurons in one ensemble tends to be small.
Such difference between membership weights leads to the difference between dendrograms. Namely, 
hierarchical structure of dendrogram can be controlled by transition parameter $\alpha^{(t)}$.
We also depict the dendrogram Y by the result of BIM, where the hierarchical structure is not similar to X.
In addition, the result of BIM cannot be controlled by tuning parameter like $\alpha^{(t)}$.
This implies that BIM may not be able to infer functional ensembles appropriately in some cases. We think that
this fact also supports the advantage of our proposed method.

\section*{Conclusion}
\label{chap:conclusion}

In this study, we extended the inference model for functional neuronal ensemble to be applicable 
regardless of data format and stationarity.
The purpose of our proposed method is to classify neurons into functional ensembles in large-scale activity data 
acquired by experimental method such as calcium imaging. For this reason, no restriction on the format of the activity or no assumption of stationarity due to physiological experimental conditions is desirable. Therefore, our proposed method without restriction on format or stationarity assumption will work effectively and can be widely used as a method of data preprocessing.
By applying our proposed method to identical synthetic data multiple times, we confirmed that it converges to a reasonable and sufficiently stable solution of functional ensembles, although our method has dependence on initial condition in MCMC. In addition, by comparing our method with BIM, we believe that our method is more useful to obtain functional ensembles and their mutual relation by adjusting the transition parameter $\alpha^{(t)}$.

The inference for functional neuronal ensembles in our study or BIM can be considered as
 preprocessing for clarifying geometry of functional neuronal network, 
 whose target size is recently increasing \cite{diana2019bayesian}. 
Therefore, functional network inference with the aid of information of functional ensemble is a future topic of our study.
As existing methods of neuronal network inference, the method using spin-glass model, which is to describe the ordered states of magnetic materials with impurities in the field of statistical physics \cite{mezard2011exact,roudi2011mean,terada2020inferring}, 
and the method of graph analysis for graphical representation of similarity 
between neurons  \cite{avitan2017spontaneous, yin2020gaussian, https://doi.org/10.48550/arxiv.2209.04117} are known for example. However, stationarity of network structure is assumed in many network inference methods. 
In addition, input neuronal activity is often limited to binary in these methods, and they cannot be applied to continuous-valued data 
such as fluorescence intensity.
Therefore, they are not sufficient as models to express functional connections between neurons with non-stationarity. 
For the use of our proposed method as preprocessing of network inference, the first issue to be considered is to generalize the network inference method without restriction of data format or stationarity assumption of input activity.

\section*{Acknowledgments}
This work is supported by KAKENHI Nos. 18K11175, 19K12178, 20H05774, 20H05776, and 23K10978.

\bibliographystyle{unsrt}
\bibliography{article}

\begin{thebibliography}{10}

\bibitem{ota2021fast}
Keisuke Ota, Yasuhiro Oisi, Takayuki Suzuki, Muneki Ikeda, Yoshiki Ito, Tsubasa
  Ito, Hiroyuki Uwamori, Kenta Kobayashi, Midori Kobayashi, Maya Odagawa,
  et~al.
\newblock Fast, cell-resolution, contiguous-wide two-photon imaging to reveal
  functional network architectures across multi-modal cortical areas.
\newblock {\em Neuron}, 109(11):1810--1824, 2021.

\bibitem{yu2021diesel2p}
Che-Hang Yu, Jeffrey~N. Stirman, Yiyi Yu, Riichiro Hira, and Spencer~L. Smith.
\newblock Diesel2p mesoscope with dual independent scan engines for flexible
  capture of dynamics in distributed neural circuitry.
\newblock {\em Nature Communications}, 12(1):1--8, 2021.

\bibitem{mezard2011exact}
Marc M{\'e}zard and J~Sakellariou.
\newblock Exact mean-field inference in asymmetric kinetic ising systems.
\newblock {\em Journal of Statistical Mechanics: Theory and Experiment},
  2011(07):L07001, 2011.

\bibitem{roudi2011mean}
Yasser Roudi and John Hertz.
\newblock Mean field theory for nonequilibrium network reconstruction.
\newblock {\em Physical Review Letters}, 106(4):048702, 2011.

\bibitem{terada2020inferring}
Yu~Terada, Tomoyuki Obuchi, Takuya Isomura, and Yoshiyuki Kabashima.
\newblock Inferring neuronal couplings from spiking data using a systematic
  procedure with a statistical criterion.
\newblock {\em Neural Computation}, 32(11):2187--2211, 2020.

\bibitem{lopes2011neuronal}
V{\'\i}tor Lopes-dos Santos, Sergio Conde-Ocazionez, Miguel~AL Nicolelis,
  Sidarta~T Ribeiro, and Adriano~BL Tort.
\newblock Neuronal assembly detection and cell membership specification by
  principal component analysis.
\newblock {\em PLOS ONE}, 6(6):e20996, 2011.

\bibitem{lopes2013detecting}
V{\'\i}tor Lopes-dos Santos, Sidarta Ribeiro, and Adriano~BL Tort.
\newblock Detecting cell assemblies in large neuronal populations.
\newblock {\em Journal of Neuroscience Methods}, 220(2):149--166, 2013.

\bibitem{molter2018detecting}
Jan M{\"o}lter, Lilach Avitan, and Geoffrey~J Goodhill.
\newblock Detecting neural assemblies in calcium imaging data.
\newblock {\em BMC Biology}, 16(1):1--20, 2018.

\bibitem{gal2021role}
Eyal Gal, Oren Amsalem, Alon Schindel, Michael London, Felix Sch\"{u}rmann,
  Henry Markram, and Idan Segev.
\newblock The role of hub neurons in modulating cortical dynamics.
\newblock {\em Frontiers in Neural Circuits}, 15, 2021.

\bibitem{diana2019bayesian}
Giovanni Diana, Thomas~TJ Sainsbury, and Martin~P Meyer.
\newblock Bayesian inference of neuronal assemblies.
\newblock {\em PLOS Computational Biology}, 15(10):e1007481, 2019.

\bibitem{kimura2021improved}
Shun Kimura, Keisuke Ota, and Koujin Takeda.
\newblock Improved neuronal ensemble inference with generative model and mcmc.
\newblock {\em Journal of Statistical Mechanics: Theory and Experiment},
  2021(6):063501, 2021.

\bibitem{miyamoto2016top}
Daisuke Miyamoto, Daichi Hirai, CCA Fung, Ayumu Inutsuka, Maya Odagawa,
  Takayuki Suzuki, Roman Boehringer, Chinnakkaruppan Adaikkan, Chie Matsubara,
  Norio Matsuki, et~al.
\newblock Top-down cortical input during nrem sleep consolidates perceptual
  memory.
\newblock {\em Science}, 352(6291):1315--1318, 2016.

\bibitem{manita2015top}
Satoshi Manita, Takayuki Suzuki, Chihiro Homma, Takashi Matsumoto, Maya
  Odagawa, Kazuyuki Yamada, Keisuke Ota, Chie Matsubara, Ayumu Inutsuka,
  Masaaki Sato, et~al.
\newblock A top-down cortical circuit for accurate sensory perception.
\newblock {\em Neuron}, 86(5):1304--1316, 2015.

\bibitem{gerstner2014neuronal}
Wulfram Gerstner, Werner~M Kistler, Richard Naud, and Liam Paninski.
\newblock {\em Neuronal dynamics: From single neurons to networks and models of
  cognition}.
\newblock Cambridge University Press, 2014.

\bibitem{neal2000markov}
Radford~M Neal.
\newblock Markov chain sampling methods for dirichlet process mixture models.
\newblock {\em Journal of Computational and Graphical Statistics},
  9(2):249--265, 2000.

\bibitem{jain2004split}
Sonia Jain and Radford~M Neal.
\newblock A split-merge markov chain monte carlo procedure for the dirichlet
  process mixture model.
\newblock {\em Journal of Computational and Graphical Statistics},
  13(1):158--182, 2004.

\bibitem{Ca2indicator}
Masamichi Ohkura, Takuya Sasaki, Junko Sadakari, Keiko Gengyo-Ando, Yuko
  Kagawa-Nagamura, Chiaki Kobayashi, Yuji Ikegaya, and Junichi Nakai.
\newblock Genetically encoded green fluorescent ${\rm ca}^{2+}$ indicators with
  improved detectability for neuronal ${\rm ca}^{2+}$ signals.
\newblock {\em PLoS ONE}, 7(12):e51286, 2012.

\bibitem{avitan2017spontaneous}
Lilach Avitan, Zac Pujic, Jan M{\"o}lter, Matthew Van De~Poll, Biao Sun,
  Haotian Teng, Rumelo Amor, Ethan~K Scott, and Geoffrey~J Goodhill.
\newblock Spontaneous activity in the zebrafish tectum reorganizes over
  development and is influenced by visual experience.
\newblock {\em Current Biology}, 27(16):2407--2419, 2017.

\bibitem{yin2020gaussian}
Hang Yin, Xinyue Liu, and Xiangnan Kong.
\newblock Gaussian mixture graphical lasso with application to edge detection
  in brain networks.
\newblock In {\em 2020 IEEE International Conference on Big Data (Big Data)},
  pages 1430--1435. IEEE, 2020.

\bibitem{https://doi.org/10.48550/arxiv.2209.04117}
Owen Forbes, Edgar Santos-Fernandez, Paul Pao-Yen Wu, Hong-Bo Xie, Paul~E.
  Schwenn, Jim Lagopoulos, Lia Mills, Dashiell~D. Sacks, Daniel~F. Hermens, and
  Kerrie Mengersen.
\newblock clusterbma: Bayesian model averaging for clustering.
\newblock {\em arXiv preprint arXiv:2209.04117v1}, 2022.

\end{thebibliography}

\end{document}